\documentstyle[12pt,epsfig]{article}

\catcode`\@=11
\def\@citex[#1]#2{%
\if@filesw \immediate \write \@auxout {\string \citation {#2}}\fi
\@tempcntb\m@ne \let\@h@ld\relax \def\@citea{}%
\@cite{%
  \@for \@citeb:=#2\do {%
    \@ifundefined {b@\@citeb}%
      {\@h@ld\@citea\@tempcntb\m@ne{\bf ?}%
      \@warning {Citation `\@citeb ' on page \thepage \space undefined}}%
      {\@tempcnta\@tempcntb \advance\@tempcnta\@ne%
      \@tempcntb\number\csname b@\@citeb \endcsname \relax%
      \ifnum\@tempcnta=\@tempcntb 
        \ifx\@h@ld\relax%
          \edef \@h@ld{\@citea\csname b@\@citeb\endcsname}%
        \else%
          \edef\@h@ld{\ifmmode{-}\else--\fi\csname b@\@citeb\endcsname}%
        \fi%
      \else
        \@h@ld\@citea\csname b@\@citeb \endcsname%
        \let\@h@ld\relax%
      \fi}%
    \def\@citea{,\penalty\@highpenalty\,}%
  }\@h@ld
}{#1}}

\def\@citeb#1#2{{[#1]\if@tempswa , #2\fi}}
\def\@citeu#1#2{{$^{#1}$\if@tempswa , #2\fi }}
\def\@citep#1#2{{#1\if@tempswa , #2\fi}}

\def\bcites{         
        \catcode`\@=11
        \let\@cite=\@citeb
        \catcode`\@=12
}

\def\upcites{         
        \catcode`\@=11
        \let\@cite=\@citeu
        \catcode`\@=12
}

\def\plaincites{      
        \catcode`\@=11
        \let\@cite=\@citep
        \catcode`\@=12
}

\newcount\hour
\newcount\minute
\newtoks\amorpm
\hour=\time\divide\hour by 60
\minute=\time{\multiply\hour by 60 \global\advance\minute by-\hour}
\edef\standardtime{{\ifnum\hour<12 \global\amorpm={am}%
        \else\global\amorpm={pm}\advance\hour by-12 \fi
        \ifnum\hour=0 \hour=12 \fi
        \number\hour:\ifnum\minute<10 0\fi\number\minute\the\amorpm}}
\edef\militarytime{\number\hour:\ifnum\minute<10 0\fi\number\minute}

\def\draftlabel#1{{\@bsphack\if@filesw {\let\thepage\relax
   \xdef\@gtempa{\write\@auxout{\string
      \newlabel{#1}{{\@currentlabel}{\thepage}}}}}\@gtempa
   \if@nobreak \ifvmode\nobreak\fi\fi\fi\@esphack}
        \gdef\@eqnlabel{#1}}
\def\@eqnlabel{}
\def\@vacuum{}
\def\marginnote#1{}
\def\draftmarginnote#1{\marginpar{\raggedright\scriptsize\tt#1}}
\def\draft{
        \pagestyle{plain}
        \overfullrule=2pt
        \oddsidemargin -.5truein
        \def\@oddhead{\sl \phantom{\today\quad\militarytime} \hfil
        \smash{\Large\sl DRAFT} \hfil \today\quad\militarytime}
        \let\@evenhead\@oddhead
        \let\label=\draftlabel
        \let\marginnote=\draftmarginnote
        \def\ps@empty{\let\@mkboth\@gobbletwo
        \def\@oddfoot{\hfil \smash{\Large\sl DRAFT} \hfil}
        \let\@evenfoot\@oddhead}
        \def\@eqnnum{(\theequation)\rlap{\kern\marginparsep\tt\@eqnlabel}%
        \global\let\@eqnlabel\@vacuum}  }

\def\blackfonts{
        \font\blackboard=msbm10 scaled\magstep1
        \font\blackboards=msbm8
        \font\blackboardss=msbm6
}

\def\nblack{            
        \def\ZZ{{Z \n{10} Z}}
        \def\NN{{N \n{14} N}}
        \def\CC{{C \n{11} C}}
        \def\RR{{R \n{11} R}}
        \def\QQ{{Q \n{12} Q}}
        \def\PP{{P \n{11} P}}
}

\def\prep{         
        \catcode`\@=11
        \input art10.sty
        \catcode`\@=12
        
        \let\small\null
        \def\blackfonts{
                \font\blackboard=msbm10
                \font\blackboards=msbm7
                \font\blackboardss=msbm5
        }
        \let\sl\it
        \twocolumn
        \sloppy
        \voffset=-2.54truecm
        \hoffset=-2.54truecm
        \flushbottom
        \parindent 1em
        \leftmargini 2em
        \leftmarginv .5em
        \leftmarginvi .5em
        \marginparwidth 48pt
        \marginparsep 10pt
        \setlength{\columnsep}{2truecm}
        \setlength{\textwidth}{25.4truecm}
        \setlength{\textheight}{17truecm}
        \baselineskip=16pt
        \oddsidemargin .18truein
        \evensidemargin .17truein
}

\def\eqalign#1{\null\,\vcenter{\openup\jot\m@th
  \ialign{\strut\hfil$\displaystyle{##}$&$\displaystyle{{}##}$\hfil
      \crcr#1\crcr}}\,}
\def\eqalignno#1{\displ@y \tabskip\centering
  \halign to\displaywidth{\hfil$\@lign\displaystyle{##}$\tabskip\z@skip
    &$\@lign\displaystyle{{}##}$\hfil\tabskip\centering
    &\llap{$\@lign##$}\tabskip\z@skip\crcr
    #1\crcr}}

\def\section{\@startsection {section}{1}{\z@}{3.ex plus 1ex minus
 .2ex}{2.ex plus .2ex}{\large\bf}}
\def\subsection{\@startsection{subsection}{2}{\z@}{2.75ex plus 1ex minus
 .2ex}{1.5ex plus .2ex}{\bf}}

\def\appendix{{\newpage\section*{Appendix}}\let\appendix\section%
        {\setcounter{section}{0}
        \gdef\thesection{\Alph{section}}}\section}

\def\abstract{\if@twocolumn
\section*{Abstract}
\else 
\begin{center}
{\bf Abstract\vspace{-.5em}\vspace{0pt}}
\end{center}
\quotation
\fi}

\catcode`\@=12

\def\Kahler{K\"ahler}
\def\kahler{k\"ahler}

\newcommand{\beq}{\begin{equation}}
\newcommand{\eeq}{\end{equation}}
\newcommand{\beqa}{\begin{eqnarray}}
\newcommand{\eeqa}{\end{eqnarray}}

\newcommand{\Z}{{\bf Z}}

\newcommand{\C}{{\bf C}}

\newcommand{\tilQ}{\widetilde{Q}}

\def\noj#1,#2,{{\bf #1} (19#2)\ }
\def\jou#1,#2,#3,{{\sl #1\/ }{\bf #2} (19#3)\ }
\def\ann#1,#2,{{\sl Ann.\ Physics\/ }{\bf #1} (19#2)\ }
\def\cmp#1,#2,{{\sl Comm.\ Math.\ Phys.\/ }{\bf #1} (19#2)\ }
\def\ma#1,#2,{{\sl Math.\ Ann.\/ }{\bf #1} (19#2)\ }
\def\ng#1,#2,{{\sl Nagoya.\ Math.\ J.\/ }{\bf #1} (19#2)\ }
\def\jd#1,#2,{{\sl J.\ Diff.\ Geom.\/ }{\bf #1} (19#2)\ }
\def\invm#1,#2,{{\sl Invent.\ Math.\/ }{\bf #1} (19#2)\ }
\def\cq#1,#2,{{\sl Class.\ Quantum Grav.\/ }{\bf #1} (19#2)\ }
\def\cqg#1,#2,{{\sl Class.\ Quantum Grav.\/ }{\bf #1} (19#2)\ }
\def\ijmp#1,#2,{{\sl Int.\ J.\ Mod.\ Phys.\/ }{\bf A#1} (19#2)\ }
\def\jmphy#1,#2,{{\sl J.\ Geom.\ Phys.\/ }{\bf #1} (19#2)\ }
\def\jams#1,#2,{{\sl J.\ Amer.\ Math.\ Soc.\/ }{\bf #1} (19#2)\ }
\def\grg#1,#2,{{\sl Gen.\ Rel.\ Grav.\/ }{\bf #1} (19#2)\ }
\def\mpl#1,#2,{{\sl Mod.\ Phys.\ Lett.\/ }{\bf A#1} (19#2)\ }
\def\nc#1,#2,{{\sl Nuovo Cim.\/ }{\bf #1} (19#2)\ }
\def\np#1,#2,{{\sl Nucl.\ Phys.\/ }{\bf B#1} (19#2)\ }
\def\pl#1,#2,{{\sl Phys.\ Lett.\/ }{\bf #1B} (19#2)\ }
\def\pla#1,#2,{{\sl Phys.\ Lett.\/ }{\bf #1A} (19#2)\ }
\def\pr#1,#2,{{\sl Phys.\ Rev.\/ }{\bf #1} (19#2)\ }
\def\prd#1,#2,{{\sl Phys.\ Rev.\/ }{\bf D#1} (19#2)\ }
\def\prl#1,#2,{{\sl Phys.\ Rev.\ Lett.\/ }{\bf #1} (19#2)\ }
\def\prp#1,#2,{{\sl Phys.\ Rept.\/ }{\bf #1C} (19#2)\ }
\def\ptp#1,#2,{{\sl Prog.\ Theor.\ Phys.\/ }{\bf #1} (19#2)\ }
\def\ptpsup#1,#2,{{\sl Prog.\ Theor.\ Phys.\/ Suppl.\/ }{\bf #1} (19#2)\ }
\def\rmp#1,#2,{{\sl Rev.\ Mod.\ Phys.\/ }{\bf #1} (19#2)\ }
\def\yadfiz#1,#2,#3[#4,#5]{{\sl Yad.\ Fiz.\/ }{\bf #1} (19#2) #3%
\ [{\sl Sov.\ J.\ Nucl.\ Phys.\/ }{\bf #4} (19#2) #5]}
\def\zh#1,#2,#3[#4,#5]{{\sl Zh.\ Exp.\ Theor.\ Fiz.\/ }{\bf #1} (19#2) #3%
\ [{\sl Sov.\ Phys.\ JETP\/ }{\bf #4} (19#2) #5]}

\def\beq{\begin{equation}}
\def\eeq{\end{equation}}
\def\beqar{\begin{eqnarray}}
\def\eeqar{\end{eqnarray}}

\newcommand{\be}{\begin{equation}}
\newcommand{\ee}{\end{equation}}
\newcommand{\bea}{\begin{eqnarray}}
\newcommand{\eea}{\end{eqnarray}}

\def\nfrac#1#2{{\displaystyle{\vphantom1\smash{\lower.5ex\hbox{\small$#1$}}%
        \over\vphantom1\smash{\raise.25ex\hbox{\small$#2$}}}}}

\def\p#1{\mskip#1mu}
\def\n#1{\mskip-#1mu}
\def\stop{\p6.}
\def\comma{\p6,}

\def\to{\rightarrow}

\def\lae{\mathrel{\mathop{\smash{\lower .5 ex \hbox{$\stackrel<\sim$}}}}}
\def\lae{\mathrel{\mathop{\smash{\lower .5 ex \hbox{$\stackrel>\sim$}}}}}

\def\pa{\partial}

\def\l:{\mathopen{:}\,}
\def\r:{\,\mathclose{:}}

\catcode`\@=11
\def\theequation{\arabic{equation}}
\catcode`\@=12

\nblack
\bcites

\nblack

\catcode`\@=11
\def\theequation{\thesection.\arabic{equation}}
\@addtoreset{equation}{section}
\@addtoreset{footnote}{section}
\@addtoreset{footnote}{subsection}
\catcode`\@=12

\newcommand{\beqn}{\begin{equation}}
\newcommand{\eeqn}{\end{equation}}
\newcommand{\beqnarray}{\begin{eqnarray}}
\newcommand{\eeqnarray}{\end{eqnarray}}
\newcommand {\bear} [1] {\begin {array} {#1}}
\newcommand {\ear} {\end {array}}

\newcommand {\beqarn} {\begin{eqnarray*}}
\newcommand {\eeqarn} {\end{eqnarray*}}

\newcommand{\tr}{{\rm Tr}}

\begin{document}

\begin{titlepage}

\begin{center}
\today
\hfill LBNL-40666, UCB-PTH-97/42\\
\hfill                  hep-th/9708044

\vskip 1.5 cm
{\large \bf Monopole Condensation and Confining Phase of $N=1$\\
 Gauge Theories
Via M Theory Fivebrane}
\vskip 1 cm
{Jan de Boer and Yaron Oz}\\
\vskip 0.5cm
{\sl Department of Physics,
University of California at Berkeley\\
366 Le\thinspace Conte Hall, Berkeley, CA 94720-7300, U.S.A.\\
and\\
Theoretical Physics Group, Mail Stop 50A--5101\\
Ernest Orlando Lawrence Berkeley National Laboratory\\
Berkeley, CA 94720, U.S.A.\\}

\end{center}

\vskip 0.5 cm
\begin{abstract}

\end{abstract}
The fivebrane of M theory is
 used in order   to study the moduli space of vacua
of confining phase $N=1$ supersymmetric gauge theories in four dimensions.
The supersymmetric vacua correspond to the condensation
of massless monopoles and confinement of photons.
The monopole and meson vacuum expectation values are computed
using the fivebrane 
configuration. 
The comparison of the fivebrane computation and the field theory analysis
 shows that
at vacua   with a classically 
enhanced gauge group $SU(r)$ the effective superpotential obtained by
the "integrating in" method   is exact for $r=2$ but is not exact for $r > 2$.  
The fivebrane configuration corresponding to $N=1$ gauge theories 
with Landau-Ginzburg
type superpotentials is studied.
$N=1$ non-trivial fixed points are analyzed using the brane geometry.

\end{titlepage}

\section{Introduction}
In the last couple of years we have learnt how string theory, M theory
and F theory can be used in order to study 
the non perturbative dynamics of low energy supersymmetric gauge
theories in various dimensions.
The two main two techniques applied in these studies are 
geometric engineering \cite{kv,geom,sd} and brane dynamics.
In the first approach one typically compactifies string theory, M theory
or  F theory on a singular Calabi-Yau
d-fold, turns off gravity and studies the gauge theory in the 
uncompactified dimensions.
In the second approach the gauge theory is realized on the 
world volume of a brane.
In this paper we will use the latter framework.

A configuration consisting of an M theory fivebrane wrapping a Riemann surface $\Sigma$ 
 can be used to study the low energy properties of  supersymmetric 
gauge theories in four dimensions.
In \cite{witten} the structure of the Coulomb branch of $N=2$ gauge theories has been determined
using the M theory 
fivebrane\footnote{The configuration consisting of a type IIA NS 5-brane wrapping
a Riemman surface $\Sigma$ describing the Seiberg-Witten curve was  studied in
\cite{sd}.
Four-dimensional abelian gauge
theory obtained from a fivebrane wrapping $\Sigma$ was 
studied in \cite{ver}.}. The study of the 
moduli space of vacua of $N=1$ SQCD 
using a configuration with an M theory fivebrane wrapping a Riemann surface 
was first done in \cite{hoo,wittennew,tau}.
  
The four dimensional supersymmetric gauge theory obtained from
the M theory  fivebrane is a compactification
of a $(0,2)$ theory in six dimensions.
As discussed in \cite{wittennew},
the brane theory has two scales, the radius of the eleventh dimension $R$ and
a typical scale of the brane configuration $L_{brane}$.
Correspondingly, there are Kaluza-Klein modes with mass $1/R$, $1/L_{brane}$. 
The supersymmetric gauge theories that we would like to study have only
one scale $\Lambda$. 
In order to correctly obtain these four dimensional 
field theories we have to find the
the values of the parameters $R,L_{brane}$ in which the brane theory and the field 
theory agree.
This, in particular,  requires a decoupling of
the Kaluza-Klein modes  of the brane theory.
This is rather difficult and
in view of these complications it is not yet precisely clear 
which quantities in field theory can be reliably computed using 
the brane theory.
It seems, however, that the brane theory
can at least be 
used for the study and computation of holomorphic objects.

In this paper we will use the fivebrane of M theory to study the moduli space of vacua
of confining phase $N=1$ supersymmetric gauge theories in four dimensions.
The supersymmetric vacua correspond to the condensation
of massless monopoles (dyons) and confinement of photons.
At points where there are mutually non-local massless dyons the dyons vevs
 and the mass gap due to dyon condensation vanish.
 These points are candidates for non trivial $N=1$ fixed points, and we will study them
 as well.

The paper is organized as follows:

Section 2 containes a brief review of the $N=2$ moduli space of vacua.
Section 3 is devoted to a field
 theory analysis of the $N=1$ gauge theory obtained by adding
to the $N=2$ superpotential an $N=1$ perturbation of the form  
$\Delta W = \sum_{k=2}^{N_c}\mu_k \tr(\Phi^k)$\footnote{The
parameter $\mu_k$ is often denoted by $g_k/k$
in the literature.} where $\Phi$ is the scalar chiral multiplet
in the adjoint representation of the gauge group.
This perturbation lifts the non singular locus of the $N=2$ Coulomb branch.
At the singular locus there are massless monopoles that can condense due to 
the perturbation, confine the gauge fields and generate a mass gap. 
We compute the monopole vacuum expectation values.
When the number of flavours $N_f$ is greater than zero there is a  meson field  $\tilde{Q}Q$
that acquires  a vaccum expectation value, where
$Q,\tilde{Q}$ constitute a quark hypermultiplet.
We compute the vev assuming a minimal 
form of the effective superpotential obtained
by the "integrating in" method. The form of the effective
superpotential is not completely fixed by symmetries and
holomorphy. Possible additional terms are 
denoted by $W_{Delta}$, and the assumption that the minimal
superpotential is exact has been called $W_{\Delta}=0$ in
\cite{int}. 

In section 4 we construct the fivebrane configuration that describes the theory with the
superpotential perturbation $\Delta W$.
The corresponding type IIA brane configuration
in the limit when $\mu_{N_c} \rightarrow \infty$
was studied in \cite{kut1,kut2}. 
In section 5 
we use the fivebrane configuration to
compute the monopole and meson vacuum expectation values as well
as the vev of the baryon operator
for $N_f \geq N_c$. 
The comparison of the fivebrane computation and the field theory analysis shows 
a complete agreement between the dyons vevs computed from field theory and the dyons vevs
computed using the fivebrane. Furthermore,
the fivebrane provides a geometrical description of the dyons vevs.
The comparison of the meson vevs
shows that that
at vacua   with enhanced gauge group $SU(r)$ the effective superpotential obtained by
the "integrating in" method with
$W_{\Delta}=0$  is exact for $r=2$ but is not exact for $r > 2$.  
Also, the fivebrane
relation between the vev of the baryon operator and the meson vevs suggests that the baryonic
branch of the $N=2$ theory is split by the perturbation, a phenomenon that does
not exist if $N=2$ is broken to $N=1$ by a mass term for the adjoint chiral multiplet.   
In section 6 we study a particular subset of the $N=2$ Coulomb branch singular locus where
a maximal number of mutually local massless monoples become massless.
We use the fivebrane configuration  to study the Higgs branches that emanate from 
this locus and relate the meson vev to those obtained when there is
only a mass term  
for the adjoint chiral multiplet.
In section 7 we study the fivebrane configuration corresponding 
to $N=1$ gauge theories with a Landau-Ginzburg
type superpotential $\sum_l  \tr (h_l \tilde{Q} \Phi^l Q)$.
In section 8 we use the brane geometry to analyze the candidates 
for $N=1$ non-trivial fixed points where mutually non-local dyons are massless.
As an illustration, the example of $SU(3)$ with $N_f=2$ is worked
out in some more detail in section~9, and section~10 is
devoted to a discussion.

The method of using brane dynamics to study supersymmetric field theories 
in various dimensions has recently been used in many other works
\cite{1,2,3,4,ejs,5,6,7,8,9,a,10,11,12,13,14,15,16,17,18,19,20,21,22,23}.

\section{Preliminaries: $N=2$ Moduli Space of Vacua}

We consider $N=2$ supersymmetric gauge theory with gauge group $SU(N_c)$ 
and $N_f$ quark
hypermultiplets in the fundamental representation.
In terms of $N=1$ superfields the vector multiplet  consists of
a field strength chiral multiplet $W_{\alpha}$ and a scalar
chiral multiplet $\Phi$
both in the
adjoint representation of the gauge group. A quark hypermultiplet consists of
a chiral multiplet $Q$ in the $N_c$ and $\tilde{Q}$
in the $\bar{N_c}$ representation of the
gauge group.
The $N=2$ superpotential takes the form 
\be
W =  \sqrt{2} \tilde{Q}_i^a\Phi^b_aQ^i_b + \sqrt{2} m_j^i\tilde{Q}_i^aQ_a^j,
\label{W}
\ee
where $a,b=1,...,N_c; i,j=1,...,N_f$ and the quark mass matrix
$m={\rm diag}[m_1,...,m_{N_f}]$.

The classical R-symmetry group is $SU(2)_R\times U(1)_R$.
The bosons in the vector multiplet are singlets
under $SU(2)_R$  while the fermions in the vector multiplet form a doublet.
The fermions in the hypermultiplet are singlets under  
$SU(2)_R$  while the scalars in the  hypermultiplet form a doublet.
The theory is asymptotically free for $N_f < 2N_c$.
The instanton factor is proportional
to $\Lambda^{2N_c-N_f}$ where $\Lambda$ 
is the dynamically generated scale.
The $U(1)_R$ symmetry is anomalous and is broken to
$Z_{2N_c-N_f}$.

The moduli space of vacua includes the Coulomb and Higgs branches. 
The Coulomb branch  is $N_c-1$ complex dimensional and
is parametrized by the gauge
invariant order parameters 
\be
u_k = \langle\tr(\phi^k) \rangle,~~~~~~~~~k=2,...,N_c
,
\label{gi}
\ee
where $\phi$ is the scalar field in the vector multiplet.
Generically along the Coulomb branch the gauge group is broken
to $U(1)^{N_c-1}$.
The Coulomb branch structure is corrected by one loop and by instantons.
The quantum Coulomb branch parametrizes a family of genus $N_c-1$
hyperelliptic curves \cite{sw,klty,af,us,apsp}
\be
y^2 = C_{N_c}^2(v)  - \Lambda_{N=2}^{2N_c-N_f}
  \prod_{i=1}^{N_f} (v +m_i) 
,
\label{curve}
\ee 
whose period matrix $\tau_{ij}$ is the low energy gauge coupling
and
where $C_{N_c}(v)$ is a degree $N_c$ polynomial in $v$ with coefficients that depend 
on the gauge invariant order parameters $u_k$, and $m_i$ $(i=1,...,N_f)$ are the quark masses.
The polynomial $C_{N_c}(v)$ is (for $N_f<N_c$) given by
\be \label{au61}
C_{N_c}(v) = \sum_{i=0}^{N_c}s_iv^{N_c-i}
\comma
\label{C}
\ee
where 
$s_k$ and $u_k$ are related by the Newton formula
\be
ks_k + \sum_{i=1}^ks_{k-i}u_i = 0,~~~~~~~~k=1,2,\ldots,N_c
\comma
\ee
with $s_0=1,s_1=u_1=0$. 
From this one can derive the following relation (for $j \geq k$)
\be \label{au21}
\frac{\partial s_j}{\partial u_k} = -\frac{1}{k}s_{j-k}
\stop
\ee

There are two types of  Higgs branches  \cite{aps}:
The non-baryonic and the 
baryonic branches. We will discuss the massless case. 
The non-baryonic branches are classified by
an integer $r$ such that 
$1 \leq r \leq min \{[N_f/2], N_c-2 \}$. The $r$-th
non-baryonic branch has complex dimension 
$2r(N_f-r)$.
The non-baryonic
branches
emanate from submanifolds in the Coulomb branch
(of dimension $N_c-r-1$ for the $r$-th non-baryonic branch)
and constitute mixed branches.
The effective theory along  the root of the baryonic branch is $SU(r)\times U(1)
 \times U(1)^{N_c-r-1}$
with $N_f$ massless quarks charged under the first $U(1)$.
Generically there are no massless hypermultiplets charged  under the last $N_c-r-1$ $U(1)$'s.
There are special points along the root where such massless matter exists.
The curve at the $r$-th non-baryonic branch root takes the form
\be
y^2 = v^{2r}\left(C_{N_c-r}(v)^2  - \Lambda_{N=2}^{2N_c-N_f}v^{N_f-2r} \right )
\stop
\label{curvenb}
\ee

There is a single baryonic branch for $N_f \geq N_c$,
where generically the gauge
group is completely broken. Its complex dimension
is $2N_f N_c - 2(N_c^2-1)$.
The baryonic branch emanates from the origin of the Coulomb branch 
The effective theory at the root of the baryonic branch is $SU(N_f-N_c) \times U(1)^{2N_c-N_f}$
with $N_f$ massless quarks charged under the $U(1)$'s
and $2N_f-N_c$ massless singlets  charged under the $2N_c-N_f$
$U(1)'s$. This can be seen by looking at the curve at the baryonic root
\be
y^2 = v^{2N_f-2N_c}\left(v^{2N_c-N_f}  + \Lambda_{N=2}^{2N_c-N_f} \right)^2
\stop
\label{curveb}
\ee 
The complete square part of (\ref{curveb}) is the required degeneration in order to have  
 $2N_f-N_c$ massless hypermultiplets.

The Higgs branches are determined classically, however
the points where they intersect each other
and the Coulomb branch are modified quantum mechanically.

\section{Breaking $N=2$ to $N=1$}

The $N=2$ supersymmetry can be  broken to $N=1$  by adding a tree level 
superpotential perturbation $\Delta W$ to the $N=2$ superpotential (\ref{W})
\be
W =  \sqrt{2} \tr(\tilde{Q}\Phi Q) +  \sqrt{2} \tr(m\tilde{Q}Q) + \Delta W 
\comma
\label{Wp}
\ee
where
\be
\Delta W = \sum_{k=2}^{N_c}\mu_k \tr(\Phi^k)
\stop
\label{dw}
\ee

In this section we will analyze the $N=1$ field theory that is obtained
from the tree level superpotential (\ref{Wp}).

\subsection{Pure Yang-Mills Theory}

 \medskip

\bigskip
{\it Dyon Condensation}
 
We consider $N=2$ pure $SU(N_c)$ Yang-Mills theory perturbed by
the superpotential $\Delta W $ (\ref{dw}).
Near points where $N_c-1$ or less mutually local dyons are massless, each charged
with respect to a different $U(1)$, the superpotential 
describing the low energy theory is
\be
W =  \sqrt{2} \sum_{i=1}^{N_c-1}\tilde{M}_iA_iM_i + \sum_{k=2}^{N_c}\mu_k U_k 
\label{Wpp}
\comma
\ee
where $A_i$ are the chiral superfield parts of the $N_c-1$  $N=2$ $U(1)$ gauge
multiplets, $\tilde{M}_i,M_i$ are the dyon hypermultiplets and
$U_k$ are the chiral superfields representing the
operators $\tr(\Phi^k)$ in the low energy theory.
The vevs of the lowest components of $A_i,M_i, U_k$ are denoted by $a_i,m_i,u_k$ 
respectively.
Note that without matter it is not possible to have a point in the moduli
space where two or more mutually local
 massless field are charged with respect to the same  $U(1)$, since otherwise
the $SQED$ effective theory at that point will have flat directions parametrizing
a  Higgs branch emanating from that point. 
Therefore the superpotential (\ref{Wpp}) describes
correctly the  pure Yang-Mills case.

The superpotential (\ref{Wpp}) is in fact exact. To show that 
we will make use of  the holomorphy and global symmetries argument \cite{hol}.
In order to apply it we need the charges of the 
fields and parameters of the theory under $U(1)_J \subset
SU(2)_R$. The assignment of charges to  the parameters and fields which restore
  $U(1)_R$ will be
useful later. The list of charges is given by  

\be
\begin{array}{cccl}
&U(1)_R&U(1)_J&\\
A_D&2&0&\\
\tilde{M}_iM_i&0&2&\\
\mu_k&2 - 2k&2\\
U_k&2k&0\\
\Lambda_{N=2}&2&0&
\end{array}
\label{list1}
\ee

The superpotential $W$ has charge two under $U(1)_J$. This  together
with the requirement for regularity at
$\mu_k = \tilde{M}_iM_i=0$ and the holomorphy constraint restrict its form
to $W= \sum_{i=1}^{N_c-1}\tilde{M}_iM_iF_i(A_i) + \sum_{k=2}^{N_c}\mu_k G_k(A)$.
The limits of small $\mu_k$ lead to the form (\ref{W}).

The low energy vacua are obtained by imposing the vanishing D-term constraints
\be
|m_i| = |\tilde{m}_i| 
\comma
\label{D}
\eeq
and the $dW=0$ constraints
\beq
-\frac{\mu_k}{\sqrt{2}} = \sum_{i=1}^{N_c-1}\frac{\pa a_i}{\pa u_k} m_i \tilde{m}_i,
~~~~~~k=2,...,N_c
\label{vac1} 
\comma
\eeq
and 
\beq
a_im_i = a_i \tilde{m}_i = 0,~~~~~~~i=1,...,N_c-1
\stop
\label{vac} 
\eeq

At a point in the moduli space where no dyons are massless we have $a_i \neq 0,i=1,...,N_c-1$.
Therefore (\ref{vac}) implies  that 
$m_i=\tilde{m_i}=0$ and thus $\mu_k=0$ by (\ref{vac1}).
This is the moduli space of vacua of the $N=2$ theory and we see that a generic point
(non-singular point) in the moduli space is lifted by the perturbation.

We  now discuss the singular points in the moduli space.
Consider a point in the moduli space where $l$ mutually local dyons
are massless. This means that some of the one-cycles shrink to zero and
that the genus $N_c-1$ curve (\ref{curve}) degenerates to a genus $N_c-l-1$ curve.
The right hand side of (\ref{curve}) takes the form
\be
C_{N_c}^2(v)  - \Lambda_{N=2}^{2N_c} =
\prod_{i=1}^{l}(v-p_i)^2
\prod_{j=1}^{2N_c -2l}(v-q_j)
\label{pr}
\ee
with $p_i$ and $q_j$ distinct. Equations (\ref{vac}) imply that
\be 
m_i=\tilde{m}_i=0,~~~~~~ i = l+1,...,N_c-1
\comma
\ee
while $m_i, \tilde{m}_i$ for
$i=1 \ldots l$ are unconstrained. Assuming the matrix
$\partial a_i/\partial u_k$ is non-degenerate there will be a 
complex $N_c-l-1$  dimensional moduli space of $N=1$ vacua which remains
after the perturbation.

The matrix $\partial a_i/\partial u_k$
can be explicitly evaluated using the relation
\be
\frac{\partial a_i}{\partial s_k}= \oint_{\alpha_i} \frac{v^{N_c-k}dv}{y}
\label{au20}
\comma
\ee
where the RHS is the integral of a  holomorphic one form on the
curve (\ref{curve}). 
At a point
where the $l$ dyons become massless we have that the cycles
$\alpha_i \rightarrow 0, i=1 \ldots l$ and
(\ref{au20}) reduces to a contour integrals around $v=p_i, i=1 \ldots l$ with the 
result
\be \label{au22}
 \frac{\partial a_i}{\partial s_k} = \frac{p_i^{N_c-k}}
 { \prod_{j \neq i}(p_i-p_j) \prod_j (p_i-q_j)^{1/2} }.
\ee
This matrix has indeed maximal rank. 
Combining  (\ref{au21}),(\ref{vac1})  and (\ref{au22}) we find the following
relation between the parameters $\mu_k$ and the dyon vevs $m_i \tilde{m}_i$
\be \label{au23}
\frac{-\mu_k}{\sqrt{2}} = 
\sum_{i=1}^l \sum_{j=2}^{N_c} \frac{-1}{k} s_{j-k} p_i^{N_c-j}
\frac{m_i \tilde{m}_i}{ 
\prod_{l \neq i}(p_i-p_l) \prod_m (p_i-q_m)^{1/2} }.
\ee
For comparison with results obtained from the branes it will
be useful to rewrite this relation in a different form. First, we
define
\be
\omega_i = \frac{\sqrt{2} m_i \tilde{m}_i}{
\prod_{l \neq i}(p_i-p_l) \prod_m (p_i-q_m)^{1/2} }.
\label{wi}
\ee
In terms of $\omega_i$ the generating function 
$\sum_{k=2}^{N_c} k \mu_k v^{k-1}$ for the $\mu_k$ is given by
\bea
\sum_{k=2}^{N_c} k \mu_k v^{k-1} & = & \sum_{k=2}^{N_c} \sum_{i=1}^l 
\sum_{j=2}^{N_c} v^{k-1} s_{j-k} p_i^{n_c-j} \omega_i \nonumber\\{}
& = & 
\sum_{k=-\infty}^{N_c} 
 \sum_{i=1}^l 
\sum_{j=2}^{N_c} v^{k-1} s_{j-k} p_i^{n_c-j} \omega_i  + {\cal O}(v^0) \nonumber\\{}
& = &  \sum_{i=1}^l 
\sum_{j=2}^{N_c} C_{N_c}(v) v^{j-N_c-1} p_i^{N_c-j} \omega_i 
  + {\cal O}(v^0) \nonumber\\{}
& = &  \sum_{i=1}^l 
\sum_{j=- \infty}^{N_c} C_{N_c}(v) v^{j-N_c-1} p_i^{N_c-j} \omega_i 
  + {\cal O}(v^0) \nonumber\\{}
& = &  \sum_{i=1}^l  \frac{C_{N_c}(v)}{(v-p_i)} \omega_i 
  + {\cal O}(v^0).
\label{au30}
\eea

Given a set of perturbation parameters $\mu_k$ and a point in the $N=2$
moduli space of vacua
specified by the set $p_i,q_j$ of (\ref{pr}),
equations (\ref{au30}) determine whether this point remains 
as an $N=1$ vacuum after the perturbation
and determine the vevs of the dyon fields $m_i \tilde{m}_i$.
It is convenient for comparison with the brane
picture to define a polynomial $H(v)$ of degree $l-1$ through
\be
\sum_{i=1}^l \frac{\omega_i}{(v-p_i)} = \frac{2H(v)}{\prod_{i=1}^l (v-p_i)}.
\label{H}
\ee
At a given point  $p_i,q_j$, $H(v)$ determines 
uniquely the dyons vevs
\be
m_i \tilde{m}_i = \sqrt{2} H(p_i) \prod_m (p_i-q_m)^{1/2}
\stop
\label{vev}
\ee

We will see that the brane picture provides a geometrical interpretation of the 
dyon vevs (\ref{vev}).

 \medskip

\bigskip
{\it Maximal Number of Mutually Local Massless Dyons}

There are $N_c$ points in the moduli space related 
to each other by 
the action of the discrete $\Z_{2N_c}$
$R$-symmetry group (\ref{list1}),
where $N_c-1$ mutually local dyons are massless.
At these points $a_i=0,  i=1,...,N_c-1$ and the curve (\ref{curve})
degenerates to a genus zero curve. 
These points correspond to the 
$N_c$ massive vacua of  $N=1$ pure Yang-Mills theory
where the discrete $\Z_{2N_c}$
$R$-symmetry is spontaneously broken to $\Z_2$.
Equations (\ref{vac1})
can be solved for generic $\mu_k$  and these $N=1$ vacua are 
generically not lifted. However, it is difficult to see from equations
(\ref{vac1}) whether there exist values of  the parameters $\mu_k$
for which these vacua are lifted.
We will see in section 6 that the brane construction predicts that these
points are not lifted for arbitrary values of the parameters 
$\mu_k$.

\subsection{$SU(N_c)$ with $N_f$ Flavours}

Consider now the addition of $N_f$ flavours to the $SU(N_c)$ Yang-Mills theory.
In the following we will compute the dyon and meson 
vevs and discuss the structure of the Coulomb
and Higgs branches of the $N=2$ theory after the perturbation.

\subsubsection{Dyon Condensation}

As in the pure Yang-Mills case, the perurbation (\ref{dw}) 
lifts the non singular locus of the
$N=2$ Coulomb branch. The computation of the dyon vevs 
along the singular locus is similar to that of
the previous subsection.
In the following we will compute the dyon vevs at the 
roots of the non-baryonic branches.
\medskip

\bigskip
{\it Non-Baryonic Branch}

The effective theory along  the root of the non-baryonic branch is $SU(r)\times U(1)
 \times U(1)^{N_c-r-1}$
with $N_f$ massless quarks $\tilde{Q},Q$ charged under the first $U(1)$
with charge $1$.
Generically there are no massless hypermultiplets charged  under the last $N_c-r-1$ $U(1)$'s.
There are special points along the root where such massless matter exists.
Along the $r$-th  non-barynoic branch root there are $2N_c-N_f$ 
special points related to each other by
a ${\bf Z}_{2N_c-N_f}$ action and for each $U(1)_i$ factor of the effective theory at the root
there is a charged massless hypermultiplet $\tilde{M}_i,M_i$.
The superpotential describing  $\tilde{M}_i,M_i$  at these points is 
\be
W = \sqrt{2}\sum_{i=1}^{N_c-r-1}\tilde{M}_iA_iM_i  
+ \sum_{k=2}^{N_c} \mu_k  U_k 
\label{WNB}
\stop
\ee
These points are similar to the above $N_c$ points in the moduli space 
of pure Yang-Mills theory  where there is a maximal number
of mutually local massless dyons
and they are  not lifted for generic $\mu_k$.
In fact, the analysis of the non-baryonic branch roots upon perturbation
is analogous  to that of the pure Yang-Mills case.
If we vary the superpotential (\ref{WNB}) with respect to 
$\tilde{M}_i,M_i,U_k$ we get
\beq
-\frac{\mu_k}{\sqrt{2}} -\sum_{j=N_c-r+1}^{N_c} \frac{\mu_j}{\sqrt{2}}
\frac{\partial{u_j}}{\partial u_k}
= \sum_{i=1}^{N_c-r-1}\frac{\pa a_i}{\pa u_k} m_i \tilde{m}_i,
~~~~~~k=2,...,N_c-r
\label{vacnb1} 
\comma
\eeq
and
\beq
a_im_i = a_i \tilde{m}_i = 0,~~~~~~~i=1,...,N_c-r-1
\stop
\label{vacnb} 
\eeq
The vanishing D-term constraints are $|m_i| = |\tilde{m}_i|$. 
The new term on the left hand side of (\ref{vacnb1})
appears because on the non-baryonic root the $U_k$
with $k>N_c-r$ are no longer independent coordinates, but
depend on $U_k$ with $k\leq N_c-r$. 

At a point in the non-baryonic branch root 
where none of the hypermultiplets $\tilde{M}_i,M_i$ are
massless we have $a_i \neq 0,i=1,...,N_c-r-1$.
Therefore (\ref{vacnb}) implies  that 
$m_i=\tilde{m_i}=0$ and we are left with $N_c-r-1$
equations for the $N_c-1$ parameters $\mu_k$. Thus, 
there is an $r$-dimensional space of $N=1$ vacua 
after the perturbation. When we discuss the non-baryonic
roots in the sequel, we will most of the time put
$\mu_k$ with $k>N_c-r$ equal to zero, so that the 
$U_k$ that appear in the superpotential are all
independent. With this additional assumption, most
of the non-baryonic root is lifted, unless some of
the hypermultiplets $\tilde{M}_i,M_i$ are massless.

If some $\tilde{M}_i,M_i$ are massless, the computation of the dyon vevs 
at the non-baryonic branch root is
analogous
to the pure Yang-Mills case where the  
hypermultiplets $\tilde{M}_i,M_i$ correspond to the dyons in that discussion.
In order to see that note that if we define $\tilde{y} = y/v^r$ in
(\ref{curvenb}) we get
at the $r$-th non-baryonic branch root 
\be
\tilde{y}^2 = C_{N_c-r}(v)^2  - \Lambda_{N=2}^{2(N_c-r)-(N_f-2r)}v^{N_f-2r} 
\stop
\label{resc}
\ee 

A point in the $r$-th non-baryonic branch root
 where $l$ mutually local dyons
are massless means that
the genus $N_c-r-1$ curve (\ref{resc}) degenerates to a genus $N_c-r-l-1$ curve
and 
it takes the form
\be
C_{N_c-r}(v)^2  -   \Lambda_{N=2}^{2(N_c-r)-(N_f-2r)}v^{N_f-2r} =
\prod_{i=1}^{l}(v-p_i)^2
\prod_{j=1}^{2(N_c-r  -l)}(v-q_j)
\label{prnb}
\ee
with $p_i$ and $q_j$ distinct. Equations (\ref{vacnb}) imply that
\be 
m_i=\tilde{m}_i=0,~~~~~~ i = l+1,...,N_c-r-1
\comma
\ee
while $m_i, \tilde{m}_i$ for
$i=1 \ldots l$ are unconstrained.
Repeating the analysis of the pure Yang-Mills case with the relation
 \be
\frac{\partial a_i}{\partial s_k}= \oint_{\alpha_i} \frac{v^{N_c-k}dv}{\tilde{y}}
= \oint_{\alpha_i} \frac{v^{N_c-r-k}dv}{y}
\label{rel}
\comma
\ee
analogous to (\ref{au30})
we get 
\be
2 H(v) \frac{C_{N_c-r}(v)}{\prod_{i=1}^{l}(v-p_i) }
 + {\cal O}(v^0) =  \sum_{k=2}^{N_c} k \mu_k v^{k-1}
 \stop
 \label{eq1}
\ee
The function $H(v)$, the dyon vevs $m_i\tilde{m}_i$ and $\omega_i$ are
related by (\ref{wi}) and (\ref{H}).

Note that the $r=0$ case is a special case of the above analysis.
 It corresponds to the general
points in the moduli space  of vacua 
which are not at the baryonic or non-baryonic roots.

 \medskip

\bigskip
{\it Baryonic Branch}
 
The effective theory at the root of the baryonic branch is $SU(N_f-N_c) \times U(1)^{2N_c-N_f}$
with $N_f$ massless quarks $\tilde{Q},Q$ charged under the $U(1)$'s
with charge $1/(N_f-N_c)$ and $2N_c-N_f$ massless singlets $\tilde{M}_i,M_i$ 
charged under only the $i$-th
$U(1)$ with charge $-1$. The superpotential 
describing  $\tilde{M}_i,M_i$ at the root is
\be
W = -\sqrt{2}\sum_{i=1}^{2N_c-N_f} \tilde{M}_iM_i A_i
+ \sum_{k=2}^{N_c} \mu_k  U_k  
\label{WB}
\stop
\ee

For each $U(1)$ factor of the effective theory at the root
there is a charged massless hypermultiplet $\tilde{M}_i,M_i$.
Thus, the baryonic branch root is similar to the $N_c$ points in the moduli space 
of pure Yang-Mills theory where there are
massless mutually local dyons  charged with respect to the different $N_c-1$ $U(1)'s$.
Therefore the baryonic branch is not lifted for generic $\mu_k$.
We will see in section 6 that the brane construction predicts that the
baryonic branch root is not lifted for arbitrary values of the parameters 
$\mu_k$.

\subsubsection{The Meson Vevs}

In the previous subsection we studied the Coulomb branch and the roots
of the Higgs branches in the presence of the perturbation 
$\Delta W $ (\ref{dw}). However, even if the root of a Higgs
branch is not lifted by a perturbation, the structure of the Higgs branch
itself may be significantly modified. In the $N=2$ theory the Higgs branch
is determined classically and it does not receive quantum corrections.
Geometrically it is a hyper{\kahler} manifold.
After adding the $N=1$ perturbation to the superpotential the structure
of the Higgs branch 
is modified by  quantum corrections
and geometrically it is a {\kahler} manifold.

In the case when the $N=1$ perturbation is only a mass term for the 
adjoint chiral multiplet, $\Delta W=\mu_2 \tr(\Phi^2)$, the structure of the Higgs
branches is the as follows \cite{aps,hoo}: The
baryonic and non-baryonic branches remain, however the non-baryonic 
branches emanate only from the points in the Coulomb branch which are not lifted.
In addition,
the structure of the  baryonic and non-baryonic branches 
depends on the parameter $\mu_2$ and they are no longer
hyper{\kahler} manifolds.
In the presence of the more general perturbation $\Delta W $ (\ref{dw})
the structure of the Higgs branches is modified   more
significantly.

In the following we will compute the vev of the meson field
$\tilde{Q}Q$ along the singular locus of the Coulomb branch.
This vev is generated by the non perturbative dynamics of the $N=1$ theory, and was clearly
zero in the $N=2$ theory before the perturbation (\ref{dw}).
 The comparison of the vev of the meson calculated by field theory
methods with the predictions from the fivebrane of M theory, which will be done
in section 5, will provide a method to test the exactness of the 
field theory description.

\medskip

\bigskip
{\it One massless dyon}

We consider the $N=1$ 
superpotential
\be
W = \sqrt{2}\tr(\tilde{Q}\Phi Q) + 
\sqrt{2}\tr (m \tilde{Q} Q) +
\sum_{k=2}^{N_c}\mu_k \tr(\Phi^k). 
\label{wp1}
\ee
In order to determine the meson vev we follow the strategy outlined in
\cite{giveon,f1,f2,f3,f4}. We first determine a low-energy effective superpotential for
a phase with one unbroken confining $SU(2)$. From that, we determine the
meson vev. The low-energy effective superpotential is not completely
fixed by symmetries and holomorphy alone, there can be corrections to
it (denoted by $W_{\Delta}$ in \cite{int}).
We will assume in the following that  $W_{\Delta}=0$.
The agreement of the meson vev  with the result obtained from the
M theory fivebrane in section 5 will justify this assumption.

Classically, the locus with one massless dyon corresponds to the cases
where $\Phi$ after diagonalization takes the form
\be
\Phi_{\rm cl}={\rm diag}(a_1,a_1,a_3,a_4,\ldots,a_{N_c})
\comma
\ee
with all $a_j$ different. For these values of $\Phi$ there is an
unbroken $SU(2)$. This form of $\Phi$ can be derived by differentiating
the superpotential with respect to $\Phi$ and putting $Q=\tilde{Q}=0$ 
which yields
\be \label{bb2}
\sum_{k=2}^{N_c}k \mu_k \Phi^{k-1} - \frac{1}{N_c}  \sum_{k=2}^{N_c}k \mu_k \tr
(\Phi^{k-1}) = 0 \comma
\ee
where the second term appears because $\Phi$ is traceless. If these equations have
a solution then we see that each diagonal component of $\Phi$ solves
an $N_c-1^{\rm th}$ order polynomial equation. As there are $N_c$ diagonal 
components in $\Phi$ this implies $\Phi$ must have two identical components.
Also, since we assumed all $a_j$ were different, these must be in one-to-one
correspondence with the solutions of this $N_c-1^{\rm th}$ order polynomial
equation. The sum of the solutions of a polynomial equation
$a z^k + b z^{k-1} + \ldots =0$ is $-b/a$, and therefore we find in this 
case that
\be
a_1 + a_3 + a_4 + \ldots  + a_{N_c} = -\frac{(N_c-1)\mu_{N_c-1}}{N_c \mu_{N_c}}.
\ee
On the other hand, we also know that $\tr(\Phi)=0$, and combining these
two equations we see that
\be \label{defa1}
a_1 = \frac{(N_c-1)\mu_{N_c-1}}{N_c \mu_{N_c}}.
\ee
One can also express the other $a_j$ in terms of $\mu_k$ but these 
expressions will not be needed in  the
present discussion.

Next, we substitute $\Phi=\Phi_{\rm cl} + \delta \Phi$ in the $N=1$
superpotential and integrate out $\delta \Phi$. This yields 
an $SU(2)$ $N=1$ theory with $N_f$ flavors and
superpotential
\be
W = \sum_{k=2}^{N_c} \mu_k \tr(\Phi^k_{\rm cl}) + 
\sqrt{2} \tr(\tilde{Q}( a_1 + m) Q) \comma
\ee
where $a_j$ should be understood as functions of $\mu_k$. 
The scale matching relation between the scale of this $N=1$
$SU(2)$ theory and the original broken $SU(N_c)$ theory
reads \cite{kss}
\be
\Lambda_{SU(2),N_f}^{6-N_f} = (N_c \mu_{N_c})^2 \Lambda_{N=2}^{2N_c-N_f}.
\ee
Next, we integrate out the quarks to end up with a pure $SU(2)$ theory, with 
superpotential 
\be \label{ab1}
W = \sum_{k=2}^{N_c} \mu_k \tr(\Phi^k_{\rm cl}) \comma
\ee
and scale 
\be
\Lambda_{SU(2)}^6 = \Lambda_{SU(2),N_f}^{6-N_f} \det ( a_1 + m).
\ee
Pure $N=1$ $SU(2)$ gauge theory has gaugino condensation, and the
final proposal for the full exact low-energy effective superpotential
is then (\ref{ab1}) plus a term due to gaugino condensation,
\be
W = \sum_{k=2}^{N_c} \mu_k \tr(\Phi^k_{\rm cl}) \pm 2 \Lambda_{SU(2)}^3
\comma 
\label{gag}
\ee
where the $\pm$ sign reflects the vacuum degeneracy of the pure $N=1$ $SU(2)$ gauge theory.
Again, the assumption that (\ref{gag}) is exact is the assumption $W_{\Delta}=0$.
In terms of the original $N=2$ scale $W$ reads
\be
W(\mu_k,m) = \sum_{k=2}^{N_c} \mu_k \tr(\Phi^k_{\rm cl}) 
\pm 2 N_c \mu_{N_c} \det ( a_1 + m)^{1/2} \Lambda_{N=2}^{N_c-N_f/2} \comma
\ee
Taking the mass matrix $m$ diagonal, we find that
the vacuum expectation value of the meson $M_i=\tr(
\tilde{Q_i} Q_i)$
\be \label{bb8}
M_i = \frac{\partial W}{\partial m_i} = \frac{\pm}{\sqrt{2}}N_c \mu_{N_c} 
\Lambda_{N=2}^{N_c-N_f/2}
\frac{\det ( a_1 + m)^{1/2}}{(a_1+m_i)}
\ee
with $a_1$ given by (\ref{defa1}).
The $\pm$ sign corresponds to two possible values
of the gauge invariant order parameters $u_k$ (\ref{gi})
parametrizing the singularities of the same $N=2$ curve.

\medskip

\bigskip
{\it More than one massless dyon}

In the following  we will generalize the previous discussion to the case with more
than one massless dyon. In field theory, such a situation can be realized by taking
\be \label{bb1}
\Phi_{\rm cl} = (a_1^{r_1},\ldots,a_k^{r_k})
\comma
\ee
by which we mean that $\Phi_{\rm cl}$ has $r_1$ times the eigenvalue $a_1$,
$r_2$ times the eigenvalue $a_2$, etc. The unbroken gauge group will
be $SU(r_1) \times \ldots \times SU(r_k) \times U(1)^{k-1}$.
Again, the logic will be to first integrate out the adjoint superfield $\Phi$, 
and after that to integrate out the quarks. 

Integrating out the adjoint superfield proceeds in two steps. First we integrate
all components of $\Phi$ that satisfy $[\Phi_{\rm cl},\Phi]\neq 0$. The
remaining components of $\Phi$ will consist of fields transforming under the
adjoint of each $SU(r_j)$, and some neutral components that we ignore. 
In the second step we integrate out the adjoints of each of the $SU(r_j)$.

After the first step level matching gives the following relation between the
original $N=2$ scale $\Lambda_{N=2}$ and the scale $\Lambda_{1,i}$ of the
$SU(r_i)$ gauge theory with adjoint and matter
\be \label{bb3}
\Lambda_{N=2}^{2 N_c - N_f} = \Lambda_{1,i}^{2 r_i - N_f} 
\prod_{j \neq i} (a_j-a_i)^{2 r_j}.
\ee

In the second step, we integrate out the adjoint in each $SU(r_j)$. The
level matching for this step involves the mass of the adjoint, which
follows by expanding the superpotential to second order
around $\Phi_{\rm cl}$
\be
m_{{\rm adj},i} = \sum_{l=2}^{N_c} l(l-1) \mu_l a_i^{l-2}.
\ee
In order to evaluate this quantity, we use the fact that all
$a_i$ are solutions of the polynomial equation (see (\ref{bb2}))
\be
\sum_{l=2}^{N_c} l \mu_l x^{l-1} 
- \frac{1}{N_c}  \sum_{k=2}^{N_c}k \mu_k \tr
(\Phi^{k-1})=0.
\ee
Therefore, we can write
\be
\sum_{l=2}^{N_c} l \mu_l x^{l-1} 
- \frac{1}{N_c}  \sum_{k=2}^{N_c}k \mu_k \tr
(\Phi^{k-1})= \phi(x) \prod_{j=1}^k (x-a_j)
\comma
\ee
where $\phi(x)$ is some polynomial of order $N_c-1-k$.
Differentiating this identity with respect to $x$ and taking
$x=a_i$ we find that
\be \label{bb4}
m_{{\rm adj},i} = \phi(a_i) \prod_{j \neq i} (a_i-a_j).
\ee
The level matching relation between the scale $\Lambda_{i,1}$ of
the $SU(r_i)$ theory with the adjoint and the scale $\Lambda_{i,2}$
of the $SU(r_i)$ theory with the adjoint integrated out
is
\be
\Lambda_{i,2}^{3 r_i - N_f} = \Lambda_{i,1}^{2 r_i - N_f}  
m_{{\rm adj},i}^{r_i}
\comma
\ee
and using (\ref{bb3}) and (\ref{bb4}) we find
\be
\Lambda_{i,2}^{3 r_i - N_f} = 
\phi(a_i)^{r_i}
\prod_{j \neq i} (a_j-a_i)^{r_i-2 r_j}
\Lambda_{N=2}^{2 N_c - N_f}.
\ee

At this moment we have pure $N=1$ gauge theory with matter.
As in the case with only one massless dyon, the last step is
to integrate out the quarks. The scale $\Lambda_i$ of the resulting
pure $N=1$ $SU(r_i)$ gauge theory is now
\be \label{bb6}
\Lambda_i^{3 r_i} = \det(m+a_i) \Lambda_{i,2}^{3 r_i - N_f} = 
\det(m+a_i) 
\phi(a_i)^{r_i}
\prod_{j \neq i} (a_j-a_i)^{r_i-2 r_j}
\Lambda_{N=2}^{2 N_c - N_f}.
\ee

The final low-energy effective superpotential is now again the
classical result with an additional term due to
gaugino condensation,
\be
W = \sum_{k=2}^{N_c} \mu_k \tr(\Phi^k_{\rm cl}) + \sum_i r_i \omega_i \Lambda_i^3
\comma
\label{gagn}
\ee
where $\omega_i$ is an $r_i$-th root of unity.
As in the one massless dyon case,
 the assumption that (\ref{gagn}) is exact means that we are assuming that
  $W_{\Delta}=0$ in \cite{int}.
For the meson vev we get
\be \label{bb7}
\sqrt{2} M_j = \frac{\partial W}{\partial m_j} = 
\sum_i \frac{\omega_i \phi(a_i)}{m_j + a_i} \det(m+a_i)^{1/r_i} 
\prod_{j \neq i} (a_j-a_i)^{1-2 r_j/r_i}
\Lambda_{N=2}^{\frac{2 N_c - N_f}{r_i}}. 
\ee
Note that the sum over $i$ in (\ref{bb7}) extends only over those
$i$ for which $r_i>1$.
In the case of one massless dyon we had $r_1=2$ and $r_j=1$ for $j>1$
and (\ref{bb7}) reduces to (\ref{bb8}). 

\medskip

\bigskip
{\it Non-Baryonic Branch} 

The non-baryonic root is a special case of the 
previous analysis for more than one massless dyon.
At the $r$-th non-baryonic root  $\Phi_{\rm cl}$ takes the form 
\be \label{bbr}
\Phi_{\rm cl} = (a_1^{r},a_2,\ldots,a_{N_c-r})
\comma
\ee
and the classical unbroken gauge group is
$SU(r)  \times U(1)^{N_c-r-1}$.
The meson vev takes the form
\be \label{bbm}
\sqrt{2} M_j = 
 \frac{\omega_i \phi(a_1)}{m_j + a_1} \det(m+a_1)^{1/r} 
\prod_{j \neq 1} (a_j-a_1)^{1-2/r}
\Lambda_{N=2}^{\frac{2 N_c - N_f}{r}}. 
\ee

\noindent
{\bf summary:}
 The perturbation $\Delta W$ (\ref{dw}) does not lift the codimension
 one singular locus of the  $N=2$  Coulomb branch where dyons become
 massless. The condensation of dyons along the singular locus confines
 photons.
Equations (\ref{wi}),(\ref{au30}), (\ref{H}) and (\ref{eq1}) determine
the vevs of the condensed dyons.
For $N_f > 0$ the meson gets a vev. Assuming $W_{\Delta}=0$ the equations
(\ref{bb8}), (\ref{bb7}) and (\ref{bbm}) give  the meson vevs along
the singular locus of the Coulomb branch.
These results will be compared with the M theory fivebrane in section 5. 
We will see that when the unbroken gauge group is $SU(2)$ there is an agreement between the
result for the meson vev in this section and the one derived via the M theory fivebrane .
When the unbroken gauge group is $SU(r), r > 2$ 
we will find disagreement with the computation using the M theory fivebrane. 
This indicates that $W_{\Delta} \neq 0$ in this case.

\section{Brane Configuration}

\subsection{Type IIA Picture}

Consider type IIA string theory in flat space-time 
where $x^0$ denotes the time coordinate
and $x^1,\ldots,x^{9}$ denote the space coordinates.
Consider first 
the type IIA picture of the $N=2$ gauge theory.
We consider the brane configuration of \cite{witten} that
preserves eight supercharges. 
The brane configuration depicted  consists of
two NS 5-branes with worldvolume
coordinates $x^0,x^1,x^2,x^3,x^4,x^5$, $N_c$ D4 branes
suspended between them
with  worldvolume
coordinates $x^0,x^1,x^2,x^3,x^6$ and $N_f$ D6 branes with 
worldvolume
coordinates $x^0,x^1,x^2,x^3,x^7,x^8,x^9$.

The D4 brane is finite in the $x^6$ direction and we consider
the four dimensional
$N=2$ supersymmetric gauge theory
on its worldvolume coordinates  $x^0,x^1,x^2,x^3$.
The theory has an $SU(N_c)$ gauge group and $N_f$ hypermultiplets
in the fundamental representation
of the gauge group \cite{witten}. The
Higgs branch is described by D4 branes suspended between D6 branes.
The motion of a D4 brane between two D6 branes is parametrized by two
complex parameters,
the $x^7,x^8,x^9$ 
coordinates together with the gauge field component  $A_6$
in the $x^6$ coordinate. 
The  brane configuration is invariant under the rotations
in the $x^4,x^5$ and $x^7,x^8,x^9$ directions ---
$U(1)_{4,5}$ and $SU(2)_{7,8,9}$.
These are interpreted as the classical $U(1)$ and $SU(2)$
R-symmetry groups of the four-dimensional theory
on the brane worldvolume.

In the following we  will consider a  perturbations of 
the form $\Delta W$ (\ref{dw}).
The brane configuration that realizes
 an $N=1$ theory with a superpotential (\ref{Wp}) has been 
 constructed and studied in \cite{kut1,kut2} 
 \footnote{The type IIA brane configuration 
 of \cite{kut1,kut2} corresponds to large coefficients in
 $\Delta W$ (\ref{dw}). The M theory fivebrane 
 configuration that we will use does not have this restriction.} .
 It consists of
NS 5-brane with worldvolume
coordinates $x^0,x^1,x^2,x^3,x^4,x^5$, 
$N_c-1$ NS' 5-branes with worldvolume
coordinates $x^0,x^1,x^2,x^3,x^8,x^9$
$N_c$ D4 branes
suspended between them
with  worldvolume
coordinates $x^0,x^1,x^2,x^3,x^6$ and $N_f$ D6 branes with 
  worldvolume
coordinates $x^0,x^1,x^2,x^3,x^7,x^8,x^9$.

Let us introduce the complex coordinates $v=x^4+i x^5,w=x^8+ix^9$.
We will take the 
$N_c$ NS' 5-branes to stretch in the $(v,w)$ coordinates.
The minima of the superpotential (\ref{Wp}) label the separation
of the NS' branes in the $v$ direction in the construction of
\cite{kut1,kut2}, a fact that we can reproduce from our
brane configuration in the limit where we send the coefficients in
$\Delta W$ to infinity.


\subsection{Fivebrane Configuration}

It has been shown in \cite{witten} that the $N=2$ brane configuration is 
described in M theory as a (generically) smooth
fivebrane with worldvolume
$R^4 \times \Sigma$ where $\Sigma$ is the  
genus $N_c-1$ curve
that determines the structure of the
$N=2$ Coulomb branch.
Denote 
$s = (x^6 + i x^{10})/R,~~ t = exp(-s)$ 
where $x^{10}$ is the eleventh coordinate of M theory
which is compactified on a circle
of radius $R$. The curve $\Sigma$ is given by an algebraic equation
in $(v,t)$ space, which for 
$N=2$ $SU(N_c)$ SQCD with $N_f$ flavors  is given by
\be
t^2 - 2C_{N_c}(v)t  + \Lambda_{N=2}^{2N_c-N_f}
  \prod_{i=1}^{N_f} (v +m_i) = 0,
\label{cu}
\ee 
where $t$ is related to $y$ in (\ref{curve}) by
$t= y + C_{N_c}(v,u_k)$.
In the M theory configuration,  $SU(2)_{7,8,9}$
is preserved but $U(1)_{4,5}$ is broken.

Let us now construct the configuration of M theory
fivebrane that corresponds to the type IIA brane realization of
the superpotential perturbation 
$\Delta W$ (\ref{dw}).
The left NS 5-brane corresponds to the asymptotic
region  $v\to \infty, t=y\sim v^{N_c}$, the right $N_c-1$ NS' 5-branes
correspond to the asymptotic
region $v\to \infty, t\sim \Lambda_{N=2}^{2N_c-N_f}v^{N_f-N_c}$.
The boundary conditions that we will impose are 

\be
\begin{array}{ccl}
w & \rightarrow & \sum_{k=2}^{N_c} k \mu_k v^{k-1}~~~~~{\rm as}~
v \rightarrow \infty,~
t \sim \Lambda_{N=2}^{2N_c-N_f}v^{N_f-N_c}\\
w & \rightarrow & 0~~~~~~~~~{\rm as}~
v \rightarrow \infty,~
t \sim v^{N_c}\\[0.2cm]
\end{array}
\label{cond}
\stop
\ee

Alternatively, if the $N_c-1$ NS' 5-branes were located at the left and 
the NS 5-brane at the right 
the boundary conditions would read 
\beq
\begin{array}{ccl}
w & \rightarrow & \sum_{k=2}^{N_c} k \mu_k v^{k-1}~~~~~{\rm as}~
v \rightarrow \infty,~
t \sim v^{N_c}\\[0.2cm]
w & \rightarrow & 0~~~~~~~~~~~~{\rm as}~
v \rightarrow \infty,~
t \sim \Lambda_{N=2}^{2N_c-N_f}v^{N_f-N_c}
\end{array}
\label{cond1}
\eeq

An alternative argument for the boundary condition (\ref{cond}) is the following.
In the $N=2$ case the left
and right NS 5-branes are parallel, both having $w=0$,
and the motion of the D4-brane
between the two NS 5-branes is the degree of freedom corresponding to
the adjoint scalar field $\Phi$. When we  deform one of the two NS 5-branes,
such motion is no longer possible, and $N=2$ supersymmetry is broken
to $N=1$. If we nevertheless try to move the D4-brane without changing
its direction, we have to take it off one of the two NS 5-branes. This
will give rise to a potential energy that gives a mass to the adjoint
scalar. The mass is proportional to the distance in the $(8,9)$ direction
between the D4-brane
and the NS 5-brane from which it is being disconnected.  If we keep one NS 5-brane at $w=0$,
and deform the other from $w=0$ to $w\equiv w(v)$, then the mass
of the adjoint is  $w(<\Phi>)$, because we can identify
$<\Phi>$ with $v$. The mass for the adjoint one gets from the
superpotential is $W'(\Phi)$, and by matching these two we see that the
boundary conditions that we have to impose on the NS 5-brane in order
to describe $\Delta W$ is 
$w(v) \sim \Delta W'(v)=\sum_{k=2}^{N_c} k \mu_k v^{k-1}$ as 
$v\rightarrow \infty$.

In the fivebrane configuration  $SU(2)_{7,8,9}$
is broken to $U(1)_{8,9}$ if the parameter $\mu_k$ 
is assigned the $U(1)_{4,5}\times U(1)_{8,9}$
charge $(2-2k,2)$.
We list below the charges of the coordinates and parameters.
\be
\begin{array}{cccl}
&U(1)_{4,5}&U(1)_{8,9}&\\
v&2&0&\\
w&0&2&\\
t&2N_c&0&\\
x&2N_c&0&\\
\mu_k&2-2k&2\\
\Lambda_{N=2}&2&0&
\end{array}
\label{list}
\ee

Comparing (\ref{list1}) and (\ref{list}) we see that $U(1)_{45} = U(1)_R$ and $U(1)_{89}=
U(1)_J$.

We do not expect to be able to construct the brane configuration
for arbitrary values of $u_k$'s.
Consider a perturbation of the form $\sum_{k=2}^{N_c-l+1} \mu_{k} \tr(\Phi^{k})$. 
We have seen that
from the field theory point of view that such a perturbation 
lifts most of the Coulomb branch of the $N=2$ theory. The moduli space of vacua that remains
is the singular part of the $N=2$ Coulomb branch where $l$ or more mutually
local dyons become massless. 
In the M-theory picture, it is possible to construct the corresponding
brane only when  the  $(v,t)$ curve degenerates to a genus $g \leq N_c-l-1$ curve.
In order to see that, note that $w$ is a function on the $(v,t)$ curve.
The boundary conditions (\ref{cond1}) or (\ref{cond})  
mean that
that $w$ is a meromorphic function of the $(v,t)$
which has a pole of order $N_c-l$ at one point.
Such a function exists only when the $(v,t)$ curve
is equivalent to a genus $g \leq N_c-l-1$ curve.

We will now analyze in detail the possible
fivebrane configurations that satisfy the right boundary conditions.
For this purpose we make two assumptions. First, we assume that
the equation defining the $N=2$ curve remains unchanged, i.e. 
is still given by (\ref{cu}).
The second assumption is that $w$ will be a rational function of
$t$ and $v$. With these assumptions, one can classify the set
of allowed functions $w$ that satisfy the appropriate boundary
conditions, as we will now demonstrate. For simplicity, we will denote
$C_{N_c}(v,u_k)$ by $C$ and 
$\Lambda_{N=2}^{2N_c-N_f}\prod_{i=1}^{N_f}(v+m_i)$ by $G$.

First, notice that any rational function of $t$ and $v$ can,
using (\ref{cu}), be rewritten in the form
\be \label{au2}
w(t,v) = \frac{a(v) t + b(v)}{c(v) t + d(v)}.
\ee
Let us denote the two solutions of (\ref{cu}) by $t_{\pm}(v)$.
It is straightforward to determine that
\be
w(t_+(v),v)+w(t_-(v),v) = \frac{2acG+2adC+2bcC+2bd}{c^2 G + 2 cdC + d^2}
\comma
\ee
and
\be
w(t_+(v),v)-w(t_-(v),v) = \frac{2(ad-bc)S\sqrt{T}}{c^2 G +2 cdC + d^2}
\comma
\ee
where
\be
 C^2(v)-  G(v) \equiv S^2(v) T(v) 
\comma
\ee
has been decomposed so that $T(v)=\prod_j(v-b_j)$ with all $b_j$
distinct.
Except for the boundary conditions on $w$ as $v\rightarrow \infty$,
we should also require that $w$ has no poles for a finite value of
$v$ since there are no other infinite NS 5-branes in the type IIA
picture. This is equivalent to the requirement that $w(t_+(v),v) \pm
w(t_-(v),v)$ has no poles. This then implies that there
should exist polynomials $H(v),N(v)$ such that
\bea \label{au3}
\frac{2acG + 2 adC + 2bcC +2bd}{c^2 G + 2  cdC + d^2} & = & 2 N \comma \\{}
\frac{2(ad-bc)S}{c^2 G +2 cdC + d^2} & = & H
\stop
\eea
By shifting $a \rightarrow a+Nc$, $b \rightarrow b+Nd$,
(\ref{au2}) and (\ref{au3}) become
\bea \label{au4}
\nonumber
w & = & N +   \frac{a(v) t + b(v)}{c(v) t + d(v)}  \comma\\{}
\nonumber
0 & = & 2acG + 2 adC+ 2 bcC+2bd \comma \\{}
H & = & \frac{2(ad-bc)S}{c^2 G + 2 cdG + d^2} \stop
\eea
The second of these equations can be decomposed as
\be
a(cG+  dC) + b(d+  cC) = 0 \comma
\ee
which implies that 
\be \label{au5}
cG+dC=-be, \qquad d+cC=ae \comma
\ee
for some
(possibly) rational function $e$. Solving for $c,d$ and substituting
this back in the third equation in (\ref{au4}) we find that it
reduces to $e=2S/H$. If we now assume that $H$, $a$ and $c$ are
given, we can solve for $b$ and $d$ using (\ref{au5}) with
$e=2S/H$. This then provides us with the most general rational
function $w$ which does not contain any poles for finite $v$.
The result is
\be \label{au6}
w = N + \frac{at + cHST - aC}{ct - c C + aS/H} \comma
\ee
where $N,H,a,c$ are arbitrary polynomials. Quite interestingly,
$w$ is independent of $a$ and $c$. One can verify that
the difference between $w$'s with different choices of
polynomials $a$ and $c$ is in fact proportional to
$t^2 -2 C t + G$ and therefore zero. The simplest cases are
to take either $a=0$ or $c=0$. In particular, if $c=0$,
\be \label{wdef}
w  =  N + \frac{H}{S}(t-C) \stop
\ee

Having determined the most general solution for $w$ that has no poles,
our next task is to impose the boundary conditions. These can be read
of from
\be \label{au8}
w(t_{\pm}(v),v) = N \pm H\sqrt{T}.
\ee
As $v\rightarrow \infty$ and $t=t_-(v) \sim v^{N_c}$, we want that 
$w \rightarrow 0$. This completely fixes $N$:
\be \label{au9}
N(v) = [ H(v) \sqrt{T(v)} ]_+ \comma
\ee
where $[f(v)]_+$ denotes the part of $f(v)$ with non-negative powers of $v$,
in a power series expansion around $v=\infty$.

With the choice (\ref{au9})
for $N$, it is guaranteed that $w(t_-(v),v)={\cal O}(v^{-1})$.

In the other asymptotic region, $v \rightarrow \infty$ and
$t = t_+(v) \sim v^{N_f-N_c}$, $w$ behaves as
\be \label{au99}
w = [2 H(v) \sqrt{T(v)}]_+ + {\cal O}(v^{-1}).
\ee
Thus, the minimal choice $H=1$ implies that $w \sim v^{k-1}$, 
if the order of $T$ is $2k-2$. 
For all other choices of $H$, $w$ grows faster than this.
This clearly shows the relation
between the genus of the degenerate Riemann surface, and the
minimal power needed in the superpotential. 

Note that imposing the other boundary condition (\ref{cond1})
simply corresponds
to the choice $N(v)=-[H(v) \sqrt{T(v)}]_+$. 

Finally, we note that $w$ satisfies the following important equation:
\be \label{au12}
w^2 -2 N w + N^2 - T H^2 = 0 .
\ee
If $T H^2 $ is of order $2k-2$, then $N^2 - TH^2$ is at most
of order $k-2$. In particular, if the genus of the degenerate
Riemann surface is zero and we choose $H=1$, then (\ref{au12})
becomes 
\be
w^2 + (a v + b) w + c=0 \comma
\ee
for some constants $a,b,c$. Solving for $v$ yields the expression
for $v$ as a function of $w$ obtained in \cite{hoo}.

In the limit where we make the coefficients in $\Delta W$ large,
both $N$ and $H$ go to infinity. In (\ref{au12}), $N^2-TH^2$
vanishes as $\Lambda_{N=2}\rightarrow 0$, and the correct limit
is one where we send $N$ and $H$ to infinity, and $\Lambda_{N=2}$
to zero in such a way that the ration $(N^2-TH^2)/N$ has a finite
limit. In this limit, (\ref{au12}) becomes $w=(N^2-TH^2)/(2N)$,
and $w\rightarrow \infty$ whenever $N\rightarrow 0$. Thus the
locations of the NS' branes in the $v$-directions are given
by the solutions of $N(v)=0$. Except for the constant term
$N(v)$ is just the derivative of the superpotential, showing
that the locations of the NS' in the $v$-direction are
given by the solutions of $W'(v)={\rm const}$. Except for
the constant, this reproduces the picture of \cite{kut1,kut2}.

\noindent
{\bf summary}: 
The brane configuration can be constructed only at the
singular locus of the $N=2$ Coulomb branch.
At  a point in the Coulomb branch where
the  $(v,t)$ curve degenerates 
and takes the form  (\ref{pr}) the fivebrane configuration is described
by
\be
w = [H(v) \sqrt{T(v)}]_+ \pm H(v) \sqrt{T(v)}
\comma
\label{w}
\ee
or equivalently
\be
w^2 -2 N w + N^2 - T H^2 = 0,
\ee
where we decomposed $C_{N_c}(v)^2- \Lambda_{N=2}^{2N_c-N_f} \prod_{i=1}^{N_f}(v+m_i)
=S(v)^2 T(v)$, 
$S(v)=\prod_{i=1}^{l}
(v-p_i)$, $T(v)=\prod_{i=1}^{2N_c-2 l}(v-q_i)$ with all
$q_i$ distinct, $H(v)$ is a polynomial in $v$ of degree $l-1$,
and $N(v)=[H(v)\sqrt{T}(v)]_+$. 
$[f(v)]_+$ denotes the part of $f(v)$ with non-negative powers of $v$
and $\pm$ refer to the two asymptotic limits $t\rightarrow 0, \infty$.

In following sections we will determine the function $H(v)$
and derive from the  fivebrane configuration
the vevs for the dyons, meson and baryon fields.

\section{Comparison to Field Theory}
We will now study
the brane description of the  superpotential perturbation $\Delta W$ (\ref{dw}) of the $N=2$
theory, and compare the results to the field theory analysis in section 3.

\subsection{Pure Yang-Mills Theory}

Consider  a point in the $N=2$ moduli space
of vacua where  the  $(v,t)$ curve degenerates to a genus $N_c-l-1$ curve.
i.e. the curve takes the form  (\ref{pr}).
As shown in the previous section, the most general deformation
of the brane  is
(see (\ref{wdef}))
\be \label{wpure}
w = N(v) + H(v) \frac{t-C_{N_c}(v)}{\prod_{i=1}^{l}
(v-p_i)} \comma
\ee
where $H(v)$ and $N(v)$ are arbitrary polynomials of $v$.
Consider the 
deformation of the left NS 5-brane. We have to impose the boundary conditions
 (\ref{cond1}). As shown in the previous section,
 the boundary condition $w  \rightarrow 0$
as $v \rightarrow \infty$ and $t \sim \Lambda_{N=2}^{2 N_c}
v^{-N_c}$ implies that $N$ has to be given by (\ref{au9}), 
explicitly 
\be
N=\left[ H \prod_{j=1}^{2N_c-2}(v-q_j)^{1/2} \right]_+
\comma
\ee 
where $[f(v)]_+$ denotes the  part of $f(v)$ with non-negative
powers of $v$.
The second boundary condition in (\ref{cond1})
shows that as $v \rightarrow \infty$,  $t \sim v^{N_c}$,
$w$ should behave as $w  \rightarrow  \sum_{k=2}^{N_c} k \mu_k v^{k-1}$.
Thus, the relation between $H(v)$ and the
values of $\mu_k$ can be determined by expanding $w$
as given in (\ref{au99}) in powers of $v$. Using that $T=(t-C)/S$
and $t=2C_{N_c}(v) + {\cal O}(v^{-N_c})$ 
we find
\be
w = 2 H(v) \frac{C_{N_c}(v)}{\prod_{i=1}^{l}(v-p_i) }
 + {\cal O}(v^0) =  \sum_{k=2}^{N_c} k \mu_k v^{k-1}
 \comma
 \label{eq}
\ee
which determines $H(v)$ in terms of $\mu_k$.

Equations (\ref{eq}) are precisely the
field theory equations (\ref{au30}) and (\ref{H}) that determine the
$N=1$ moduli space of vacua after perturbation and the dyon vevs.
We see that the  M theory fivebrane describes correctly the fact that
only the singular part of the $N=2$ Coulomb
branch  is not lifted and reproduces the equations that 
determine the vevs of the massless dyons along the singular locus.
The geometrical interpretation of the dyon vevs (\ref{vev})
will be given at the end of section~5.2.1.

\subsection{$SU(N_c)$ with $N_f$ Flavours}

The computation of the dyons vevs along the singular locus is similar to that of
the previous subsection. Since $t=2C_{N_c}(v)+{\cal O}(v^{N_f-N_c})$, 
(\ref{eq}) is only valid if $N_f\leq N_c$. However, this does not
contradict (\ref{au30}), because that result assumes the form
of the curve (\ref{curve}) which is no longer correct for $N_f\geq N_c$.
We have not checked the case $N_f>N_c$ in detail, but believe the
correspondence between the field theory and five-brane will still
be valid, in particular the dyon vevs are still given by (\ref{vev})
where $H(v)$ is the polynomial entering the description of the five
brane geometry.

In the following we will compute the dyons vevs at the roots of the non-baryonic branches,
for the cases where $N_f-2r \leq N_c-r$.

\subsubsection{Dyon Condensation}

\bigskip
{\it Non-Baryonic Branch}

Consider  a point  at the $r$-th non-baryonic
branch root where the  $(v,\tilde{t}= \tilde{y} + C_{N_c-r}(v))$ curve 
(\ref{resc}) degenerates to a genus $N_c-r-l-1$ curve and takes the form
  (\ref{prnb}).
Using the results in section~4 in the presence of matter, the most general deformation
of the brane  in our case is
(see (\ref{wdef}))
\be \label{wpurenb}
w = N(v) + H(v) \frac{\tilde{t}-C_{N_c-r}(v)}{\prod_{i=1}^{l}
(v-p_i)} \comma
\ee
where $H(v)$ and $N(v)$ are arbitrary polynomials of $v$.
We restrict $H(v)$ to be at most of order $l-1$, as discussed in
section~3.2.1.
Consider the 
deformation of the left NS 5-brane. 
The boundary condition $w  \rightarrow 0$
as $v \rightarrow \infty$ and $\tilde{t} \sim \Lambda_{N=2}^{2 (N_c-r)}
v^{-N_c+r}$ implies again (\ref{au9}), i.e.
\be
N=\left[ H \prod_{j=1}^{2(N_c-r-1)}(v-q_j)^{1/2} \right]_+
\stop
\ee 
The second boundary condition, which says that as
$v \rightarrow \infty$,  $\tilde{t} \sim v^{N_c-r}$,
$w$ should behave as $w  \rightarrow  \sum_{k=2}^{N_c} k \mu_k v^{k-1}$,
yields again the relation between $H(v)$ and $\mu_k$ by
expanding (\ref{au99}) in powers of $v$. Using 
$\tilde{t} = 2C_{N_c-r}(v) + {\cal O}(v^{N_f-r-N_c})$ 
we obtain
\be
w = 2 H(v) \frac{C_{N_c-r}(v)}{\prod_{i=1}^{l}(v-p_i) }
 + {\cal O}(v^0) =  \sum_{k=2}^{N_c} k \mu_k v^{k-1}
 \comma
 \label{eq2}
\ee
which determines $H(v)$.

Equations (\ref{eq2}) are precisely the
field theory equations (\ref{eq1}) that determine the
structure of the non-baryonic branch
after perturbation and the dyon vevs.
We see that the fivebrane in M theory describes correctly the moduli space
of vacua of the $N=1$ theory corresponding to the non-baryonic, as well as
the $r=0$ case which
corresponds to the general
points in the moduli space  of vacua 
which are not at the baryonic or non-baryonic roots.

 \medskip

\bigskip
{\it Baryonic Branch}

At the baryonic branch root
 the $(v,t)$ curve
  degenerates at $2N_c-N_f$ points and factorizes
into two rational curves:
\beqar
C_L:~~~~~~t &=& v^{N_c},~~~~~ w=0, \nonumber\\ 
C_R:~~~~~~t &=& \Lambda_{N=2}^{2N_c-N_f}v^{N_f-N_c},~~~~~ w=0
\stop
\eeqar
This makes the construction of the fivebrane configuration easy in this case.
In order to satisfy the boundary conditions that describe the
$N_c-1$ left NS' 5-branes we
simply replace $t=v^{N_c}, w=0$ by
$ t=v^{N_c},      w=  \sum_{k=2}^{N_c} k \mu_k v^{k-1}$ in order to satisfy the 
boundary conditions. The fivebrane curve factorizes
into two rational curves:
\beqar
\tilde{C}_L:~~~~~~t &=& v^{N_c},~~~~~ w=\sum_{k=2}^{N_c} k \mu_k v^{k-1}
 \nonumber\\ 
C_R:~~~~~~t &=& \Lambda_{N=2}^{2N_c-N_f}v^{N_f-N_c},~~~~~ w=0
\stop
\label{bc}
\eeqar

The field theory analysis suggested that the baryonic branch
root is not lifted for generic $\mu_k$.
We see from 
the brane picture that the baryonic branch
root is  not lifted for arbitrary values of the parameters $\mu_k$.

 \medskip

\bigskip
{\it Geometrical Interpretation of the Dyon Vevs}

The dyon vev (\ref{vev}) $m_i\tilde{m}_i$ is equal to 
$\sqrt{2} (H\sqrt{T})(p_i)$. This is equal (up to a factor
of $\sqrt{2}$) to the difference
between the two finite values of $w$ (\ref{w}) as we take $v=p_i$. 
The singular $N=2$ curve (\ref{pr}), (\ref{cu})
has a double point at $v=p_i,t=C_{N_c}(p_i)$, caused by the
existence of a massless dyon.
After the perturbation $\Delta W$ (\ref{dw}) this double point splits into two
separate points in $(v,t,w)$ space, and the distance between the points
in the $w$ direction is exactly the dyon vev
of the dyon that became massless at this point in the $N=2$
theory. Thus, giving a vev to the field that was responsible for the
singularity resolves the singularity. This is analogous to the resolution
of the conifold singularity \cite{gms}, where black holes rather than
dyons cause and resolve the singularity. 
This provides a simple geometrical interpretation of
the dyon vevs in the brane picture. In the case of Argyres-Douglas
points (see section~8), the singularity is worse than a double point, 
the dyon vev vanishes at the singularity
and the singularity is not completely resolved.

 \begin{figure}[htb]
\begin{center}
\epsfxsize=5in\leavevmode\epsfbox{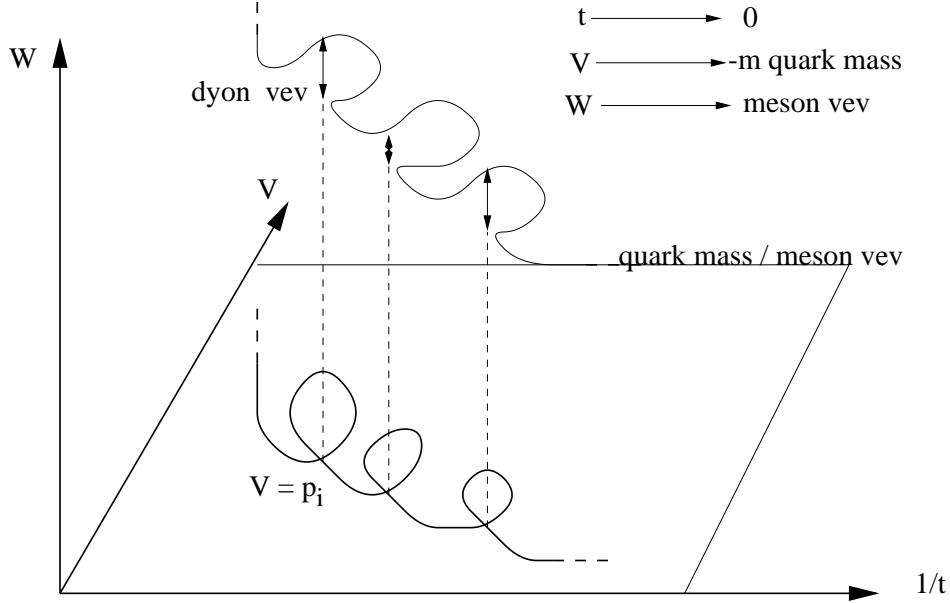}
\end{center}
\caption{The Dyon and Meson Vevs in the Brane Geometry}
\label{fig}
\end{figure}

\subsubsection{The Meson Vevs}

In \cite{hoo} the eigenvalues of the meson matrix were identified
with the  values of $w$ at $t=0, v=-m_i$. 
One way to see this identification is to note
that in  the Type IIA set-up in which all the D6-branes are sent to
the infinity $x^6=+\infty$,
there are $N_f$ semi-infinite D4-branes ending on the right NS 5-brane
from the right. The values of $w$
at $t=v=-m_i$ 
are the asymptotic positions in the $w=x^8+ix^9$ direction of these
semi-infinite D4-branes.
Moreover, the order of the zero of $t$ 
at these values of $w$ is the 
the number of D4-branes at these values
in the limit $x^6\to +\infty$.

The $U(N_f)$ symmetry  associated with these $N_f$
 semi-infinite D4-branes is a global symmetry
of the four-dimensional field theory
on the D4-branes which are finite in the $x^6$ direction.
When the D4-branes are separated from each other in the $w$ direction
the global symmetry is broken.
The only quantity 
with $U(1)_{8,9}$ charge 2 that can break the $U(N_f)$ flavor
symmetry
 is the meson vev $M^i_j=\tilQ^iQ_j$.

\bigskip
{\it One massless dyon}

 \medskip

Let us now compute the finite values of $w$ at $t=0, v=-m_i$ and
compare to the meson vevs.
 
At the locus where there is
one massless dyon we have that
\be \label{ab3}
C_{N_c}^2(v) - \Lambda_{N=2}^{2 N_c- N_f} \prod_{i=1}^{N_f} (v+m_i) = 
(v-p)^2 T(v).
\ee
and the function $w$ is given by 
\be
w = [H \sqrt{T(v)}]_+ \pm H \sqrt{T(v)}
\comma
\ee
where $H$ is a constant. We assume $N_f<N_c$. Then 
we have the following important simplification
\be \sqrt{T(v)} = \frac{C_{N_c}(v)}{v-p} + {\cal O}(v^{-1}).
\ee
We can always decompose 
\be \label{ab4}
C_{N_c}(v) = C_{N_c}(p) + (v-p)\tilde{C}_{N_c}(v)
\comma
\ee
for some $N_c-1$ degree polynomial $\tilde{C}_{N_c}$, and we
see that 
\be \label{ab2}
\sqrt{T(v)} = \tilde{C}_{N_c}(v) + {\cal O}(v^{-1}) \quad \longrightarrow
\quad  [\sqrt{T(v)}] = \tilde{C}_{N_c}(v).
\ee
We are interested in the finite values of $w$ as $v\rightarrow
 -m_i$ and $y \rightarrow 0$. Using (\ref{ab2}) we find
\be \label{ab5}
w_i = w(v \rightarrow -m_i) = 
H \tilde{C}_{N_c} (-m_i)  \pm H \sqrt{T(-m_i)}.
\ee
Using (\ref{ab3}) it is clear that $\sqrt{T(-m_i)} = C_{N_c}(-m_i)/(-p-m_i)$
and using the decomposition (\ref{ab4}) we find that
\be
\sqrt{T(-m_i)} =  \frac{C_{N_c}(p)}{-p-m_i} + \tilde{C}_{N_c}(-m_i).
\ee
Inserting this back in (\ref{ab5}), and taking the minus rather than the
plus sign (which is the sign that corresponds to $t \rightarrow 0$) we
get 
\be \label{bb9}
w_i = H \frac{C_{N_c}(p)}{p+m_i}
\comma
\ee
To work out $C_{N_c}(p)$ we insert $v=p$ in (\ref{ab3}) and from that obtain
\be
w_i = H \Lambda_{N=2}^{ N_c- N_f/2} \frac{\det ( p + m)^{1/2}}{(p+m_i)}.
\ee
The last thing we have to do is to compute $p$ and $H$. The asymptotic behavior
of $w$ for large $v$ is 
\be \label{518}
w \sim 2 H \frac{C_{N_c}(v)}{v-p} \sim 2 H v^{N_c-1} + 2 H p v^{N_c-2} + \ldots
\comma
\ee
and this should be equal to $N_c \mu_{N_c} v^{N_c-1} + (N_c-1) \mu_{N_c-1} v^{N_c-2} 
+ \ldots$ from which we derive that 
\be
2H = N_c \mu_{N_c} , \qquad p =\frac{(N_c-1) \mu_{N_c-1}}{N_c \mu_{N_c}}
\comma
\ee
which then finally gives us for the finite value of $w$
\be
w_i = \frac{1}{2} N_c \mu_{N_c} 
\Lambda_{N=2}^{N_c-N_f/2}
\frac{\det ( a_1 + m)^{1/2}}{(a_1+m_i)}
\stop
\label{Wi}
\ee
Comparing (\ref{Wi}) and (\ref{bb8}) we see that
up to a factor of $\sqrt{2}$,
the values of $w$ at $t=0,v=-m_i$ are  exactly the meson vevs  derived from
field theory assuming the low energy effective superpotential (\ref{gag}), which was
derived using the "integrating in" method with  $W_{\Delta}=0$.

\medskip

\bigskip
{\it More than one massless dyon}

Let us now compare the result (\ref{bb7}) for the case with more than
one massless dyon to the finite values of $w$.  If there
are $l$ massless dyons the $N=2$ curve can be factorized as
\be \label{bb10}
C_{N_c}(v)^2 - \Lambda_{N=2}^{2N_c-N_f} \prod_{i=1}^{N_f} (v+m_i) =
\prod_{i=1}^l (v-p_i)^2 T(v)
\ee
In this case, $w$ is given by
\be w=[H(v)\sqrt{T(v)}] \pm H(v)\sqrt{T(v)},
\ee
where $H(v)$ is a polynomial of degree $l-1$. 
As in the case with one massless dyon, for $N_f \leq N_c$
we can write
\be
H(v) \sqrt{T(v)} = \frac{H(v)C(v)}{\prod_{i=1}^l (v-p_i)}
+ {\cal O}(v^{-1}).
\ee
We can always decompose
\be \label{bb11}
H(v) C(v) = G_0(v) + G_1(v) \prod_{i=1}^l (v-p_i)
\ee
with $G_0(v)$ a polynomial of order $l-1$. 
The same derivation that led to (\ref{bb9}) shows that in this
case the $j$-th finite value of $w$ is equal to
\be
w_j = \frac{G_0(-m_j)}{\prod_{i=1}^l (m_j+p_i)}.
\ee
It remains to determine $G_0(-m_j)$. Taking $v=p_i$ in
(\ref{bb11}) and (\ref{bb10}) we find
\be
G_0(p_i) = H(p_i) \sqrt{
\prod_{j=1}^{N_f} (m_j+p_i)} \Lambda_{N=2}^{N_c-N_f/2}
\ee
These are $l$ linear equations for the $l$ coefficients of
$G_0$ and can be used to completely determine $G_0$ in terms
of $p_i$ and $m_j$, and also to determine $G_0(-m_j)$.
We omit the details of this calculation, but just give the
final result for $w_j$
\be \label{bb12}
w_j = \sum_{i=1}^l \frac{H(p_i) \det(m+p_i)^{1/2}}{
\prod_{t \neq i} (p_t-p_i) (m_j+p_i)} 
\Lambda_{N=2}^{N_c-N_f/2}.
\ee

Comparing (\ref{bb12}) and (\ref{bb7}) we see that if $r_1=\ldots=r_l=2$ and $r_j=1$ 
for $j>l$, then (\ref{bb7}) agrees with
(\ref{bb12}), if we identify $\phi$ with $H$ 
and  $a_i$ with $p_i$. Therefore it seems that 
$l$ massless dyons can be described by an $SU(2)^l$
confining $N=1$ theory, and the "integrating in" method works with $W_{\Delta}=0$.  

It is interesting to note that the above algebraic results continue to 
agree even when $2l > N_c$, in which case the field 
theoretical derivation is  senseless.
This may indicate that the low-energy effective superpotential
is still fine, although its derivation is not. 

When we compare (\ref{bb12}) and (\ref{bb7}) with  at least one of the $r_i$ greater
than two  we see a disagreement.
In this case there is a corresponding classical enhancement 
of the gauge group to $SU(r_i)$.
We interpret the disagreement as an indication that the effective low energy superpotential (\ref{gagn}) 
does not provide an exact description and  
 $W_{\Delta} \neq 0$ in the
"integrating in" method. Notice that (\ref{bb12}) has a well-defined
limit if we send $p_i \rightarrow p_j$, as long as $p_i+m_j \neq 0$. 
The case when $p_i+m_j=0$ is briefly commented upon in section 9.

\medskip

\bigskip
{\it Non-Baryonic Branch}

At a generic point in the  $r$-th non-baryonic root, with $l$ additional
massless dyons, the fivebrane result for the 
meson vevs can be read from (\ref{bb12})
\be \label{bb13}
w_j =  \sum_{i=1}^l \frac{H(p_i) \det(m+p_i)^{1/2}}{
\prod_{t \neq i} (p_t-p_i) (m_j+p_i)} 
\Lambda_{N=2}^{N_c-N_f/2}.
\ee
This result is the same as one would obtain for a confining $SU(2)^l$
subgroup in $SU(N_c-r)$, but does not at all resemble the result
(\ref{bbm}) for a confining $SU(r)$ subgroup in $SU(N_c)$, nor
does it look like any of the results obtained with
a confining $SU(r) \times SU(2)^l$ in $SU(N_c)$.
As we discussed previously 
this is as an indication that the effective low energy superpotential (\ref{gagn}) 
does not provide an exact description at the $r$-th non-baryonic branch root and  
$W_{\Delta} \neq 0$ in the
"integrating in" method.

\subsubsection{The Baryon Vev}

The vev of the baryon operators vanishes on the non-baryonic
branches and is non zero on the baryonic branch.
The curve at the baryonic root  consists  of two branches
(\ref{bc}).
In \cite{hoo} the vev for the baryon operator $\tilde{B}B$
when the vev of the meson
matrix vanishes 
was identified (up to chiral rotation)
with the distance between these two branches at $w=0$.
This identification   reproduced upon varying $w$, namely
giving a vev to the meson, the correct field theory equation.
In our case we do not have the field theory result.
However, we can still get the brane prediction for the corresponding equation.
We get
\beq
\Delta t = v^{N_c} - \Lambda_{N=2}^{2N_c-N_f}v^{N_f-N_c}
\comma
\eeq
where $\sum_{k=2}^{N_c}k \mu_k v^{k-1} = m$, $\Delta t = \tilde{B}B$ and
$m$ is the meson vev.

We see that  for a given vev $m$ for the meson, 
the baryon vev can take several
values. This indicates that
the baryonic branch splits in several
parts, which is a novel phenomena that we do not have when we only add
a mass for the adjoint chiral multiplet.
Another possible interpretation of this phenomenon is that we are
not using the right degrees of freedom, namely that we should use
several meson and baryon operators to describe the branch.

\section{Maximal Number of Mutually Local Massless Dyons}

In \cite{hoo} the fivebrane description of the $N=2$ theory perturbed by adding
a mass term for the adjoint chiral multiplet was studied. With such a perturbation,
most of the Coulomb branch is lifted besides a discrete set of points.
These points lie on orbits of $\Z_{2N_c-N_f}$ and correspond to 
 points in the moduli space of vacua
 with the maximal number ($N_c-1$) of mutually local massless dyons.
The root of the baryonic branch as well as the baryonic branch itself remain.
The non baryonic branches remain but instead of
being mixed branches they emanate from points.
The $r$-th non-baryonic branch that 
emanated from
a submanifold of dimension $N_c-r-1$ in the Coulomb branch
is now emanating from $2N_c-N_f$ points related
to each other by  $\Z_{2N_c-N_f}$ 
\footnote{The 
$r=N_f/2$ case ($N_f$ even) is exceptional, 
the $\Z_2$ subgroup is unbroken
and the $\Z_{2N_c-N_f}$ orbit consists of $N_c-N_f/2$ points.}.
At these points there are $N_c-r-1$  additional mutually local massless dyons.

The above points are characterized geometrically by the fact that the 
genus $N_c-1$ hyperelliptic curve degenerates to a genus zero curve. 
In this section we construct the fivebrane configuration, corresponding
to the $N=2$ theory perturbed by  the superpotential $\Delta W$ (\ref{dw}),
at these 
points.
We have $T(v)=v^2 + a v + b$ for certain $a,b$.
Following the same
arguments
as in \cite{hoo} there exists a 
coordinate $q$ such that
\bea \label{au60}
v & = & \frac{(q-q_+)(q-q_-)}{q}  \nonumber \\{}
t &  = & q^{N_c-N_f} (q-q_+)^r (q-q_-)^{N_f-r}.
\eea
the values of the parameters
$q_{\pm}$ are determined by the requirement that $v,t$ in
(\ref{au60}) satisfy (\ref{cu}) and take the form \cite{hoo}

\beqar
q_+ &=& 
-\frac{N_c-N_f+r}{N_c - r}q_- \nonumber\\
q_- &=& \left(\frac{(-1)^{r+N_c}}{4}
\left(\frac{N_c-r}{N_c - N_f + r}\right )^{N_c-N_f+r} \right )^{\frac{1}
{2N_c-N_f}}  \Lambda_{N=2}
\stop
\label{sol}
\eeqar

Consider now the superpotential deformations  $\Delta W$ (\ref{dw}).
$w$ will be a meromorphic function on  the Riemann surface, and
is therefore a function of $q$. This function can be
explicitly determined and reads
\be
w = \sum k \mu_k \left[ \left( \frac{(q-q_+)(q-q_-)}{q} \right)^{k-1} 
\right]_{>}
\comma
\label{wsol}
\ee
where now $[a(q)]_>$ denotes the terms with a positive power of $q$
in the power series expansion of $a(q)$ (i.e. we also drop the
constant piece). 
If there is only a perturbation with $\mu_2 \neq 0$, 
we find $w=2\mu_2 q$ as expected.
Using the explicit expression for $w$ as a function
of $q$ we can now determine the two finite values of $q$ as $v\rightarrow 0$,
i.e. if $q\rightarrow q_{\pm}$. We get that $w_{\pm}$ is the
constant term in the power series in $x$ of
\be \label{au50}
w_{\pm} = \left. \sum_k k \mu_k (-1)^{k} (x+q_{\pm}) 
\frac{(x+q_+)^{k-2} (x+q_-)^{k-2}}{x^{k-2}} \right|_{x^0}
\stop
\ee
When only $\mu_2 \neq 0$ we see that $w_{\pm}=2 \mu_2 q_{\pm}$
in agreement with \cite{hoo}.

\subsection{ Pure Yang-Mills}
In section 3 we 
noted that there are 
 $N_c$ points in the moduli space related 
to each other by 
the action of the discrete $\Z_{2N_c}$
$R$-symmetry group (\ref{list1}),
where $N_c-1$ mutually local dyons are massless.
It was not clear from the field theory analysis under what conditions these vacua
are not lifted. The brane picture suggests that these points are not lifted
for arbitrary values of the perturbation parameters $\mu_k$.
The fivebrane configuration at one of this points takes
the form

\beqar
v &=& t^{1/N_c} -  \Lambda_{N=2}^{2} (4t)^{-1/N_c}\nonumber\\
w &=& \sum_{k=2}^{N_c} k \mu_k [(t^{1/N_c} -  \Lambda_{N=2}^{2} (4t)^{-1/N_c})^{k-1}]_>
\comma
\eeqar
while the configurations at the other $N_c-1$ points are constructed by
applying the    
$Z_{2N_c}$ action. Again, $[f(t)]_>$ means keeping only the positive powers
of $t$. 

\subsection{$SU(N_c)$ with $N_f$ Flavours}

\medskip

\bigskip
{\it Massless Quarks}

For each $r<N_f/2$ there are $(2N_c-N_f)$ solutions (\ref{au60}), (\ref{sol})
and (\ref{wsol})  related
by the $\Z_{2N_c-N_f}$ action, while for
$r=N_f/2$ ($N_f$ even) there are $(N_c-N_f/2)$ solutions.

 The  values $w_{\pm}$ of $w$ at $t=0$ (\ref{au50}) are interpreted
 as the $(r,N_f-r)$
eigenvalues of the meson matrix 
$M^i_j=\tilQ^iQ_j$.
The $r$-th Higgs branch 
emanating from the $2N_c-N_f$ points where the meson matrix takes the 
diagonal form with these eigenvalues 
 consists of the orbits of the complexified
flavor group
$GL(N_f,\C)$ acting on this  diagonal meson matrix.
The diagonal meson matrix  is invariant under the subgroup
$GL(r,\C)\times GL(N_f-r,\C)$ which implies that
the Higgs branch  is the homogeneous space
$GL(N_f,\C)/(GL(r,\C)\times GL(N_f-r,\C))$. It has a complex dimension
$2r(N_f-r)$ which is the 
dimension of the $r$-th
non-baryonic branch.
It is interesting to note
that quations (\ref{au50}) express the eigenvalues of the meson matrix
 for a generic 
perturbation  $\Delta W$ in terms
of the meson vevs in the presence of  only a mass term for the adjoint
chiral multiplet  $\mu_2 \tr(\Phi^2)$.

\medskip

\bigskip
{\it Massive Quarks}

A similar construction of the fivebrane configuration
can be done when  in the presence of a 
quark mass term $m\sum_i \tilQ^iQ_i$. 
In this case, there is no baryonic branch and the curve $C$ does
not factorize.
The difference between the massive and massless fivebrane 
configuration 
is that  $v$ is replaced by
$v+m$ in (\ref{au60})
\be
v + m =  \frac{(q-q_+)(q-q_-)}{q}  
\comma
\ee
and the relation between $q_{\pm}$ in (\ref{sol}) is modified to
\beq
q_+ = 
-\frac{1}{N_c - r}\left((N_c-N_f+r)q_- + m \right) 
\stop
\label{solm}
\eeq
In addition, we need to modify (\ref{wsol}), by
replacing $(q-q_+)(q-q_-)/q$ by $(q-q_+)(q-q_-)/q-m$,
and (\ref{au50}) has to be modified accordingly.
For every $r=0,1,\ldots,[N_f/2]$, there are $2N_c-N_f$
solutions, which however are not related 
by the discrete $R$-symmetry group $Z_{2N_c-N_f}$
which is broken by the quark mass term.
As in the massless case, the $r$-th Higgs branch emanating from 
these points has complex dimension
$2r(N_f-r)$.

More complicated mass terms for the quarks can also be treated, but
the equation for $t$ as a function of $q$ and the equations for the
meson vevs will become much more complicated.

\section{Landau-Ginzburg Type Deformations}

$N=1$ gauge theories with a LG type superpotential 
\be \label{def2}
W =\sum_l  \tr (h_l \tilde{Q} \Phi^l Q)
\comma
\ee
have been studied in
\cite{kapu,givv}.
These theories have a Coulomb branch parametrized by the gauge invariant order
parameters constructed from the vev of $\Phi$.
The Coulomb branch parametrizes a family of  hyperelliptic curves
\be \label{cur2}
t^2 - 2 C_{N_c}(v) t + \Lambda_{N=2}^{2N_c-N_f} 
\det(m+\sum h_l v^l)=0 
\stop
\ee
The low energy 
 gauge coupling of the $N=1$ theory is the period matrix of the 
  hyperelliptic curve (\ref{cur2}).
However, the theory has only $N=1$ supersymmetry and therefore the special geometry
structure of $N=2$ theories no longer exists
and the {\Kahler} potential is not encoded in the curve.

In the following we will discuss the M theory fivebrane configurations that
correspond to the $N=1$ theories with the superpotential (\ref{def2}).
The simplest example 
that gives rise to an $N=1$ theory with a superpotential of the form (\ref{def2})
is the one obtained by rotating simultaneously the two
NS 5-branes in the IIA picture of the $N=2$ theory from $(x^4,x^5)$
to $(x^8,x^9)$
by a fixed angle.
Let us construct 
the rotated brane configuration in M theory.
The $N=2$  brane configuration can be written in the form \cite{witten}
\bea \label{u1} \nonumber
tz & = & \Lambda_{N=2}^{2N_c-N_f} \prod_{i=1}^{N_f} (v+m_i), \\{}
\nonumber
t+z & = & 2 C_{N_c}(v), \\{}
w & = & 0 
\stop
\eea
The first of these equations  describes $N_f$
 D6-branes located at $t=z=0$, $v=-m_i$.
Recall that in M theory, 
the D6-branes are Kaluza-Klein Monopoles described by
a Taub-NUT space \cite{tow}.
As a complex surface, the Taub-NUT space is the same as 
the ALE space of the $A_{n-1}$-type described by the first equation in (\ref{u1}).
The last two equations in (\ref{u1}) describe the geometry of
the Riemann surface.  When we rotate the
two NS-branes in the type IIA picture 
by an angle $\phi$ in the $(x^8,x^9)$ direction, we keep the D6 branes
fixed and therefore in the M theory description
we do not change the ALE space defining equation. 

After the rotation, the last two equations in (\ref{u1}) are modified to
\bea  \label{au40}
t+z  & = & 2 C_{N_c}(v(\phi)), \\{}
w(\phi) & = & 0 
\comma
\eea
where $v(\phi)$, $w(\phi)$ are the rotated coordinates
\be
w(\phi) = w \cos \phi - v \sin \phi, \qquad
 v(\phi) = v \cos \phi + w \sin \phi.
\ee
If we eliminate $w$ from (\ref{au40}), and introduce the new
variable $\tilde{v}=v(\phi)$, the rotated fivebrane configuration is
\bea \label{au41}
tz & = & \Lambda_{N=2}^{2N_c-N_f} \prod_{i=1}^{N_f} (\cos \phi \tilde{v} + m_i),
\\{}
t+z  & = & 2 C_{N_c}(\tilde{v}),
\\{}
w & = & \tilde{v} \sin \phi.
\eea
The first and second equation  describe the Coulomb branch
of $N=1$ SYM with a $W=\sqrt{2} \cos \phi \tr( \tilde{Q} \Phi Q)$
superpotential (\ref{cur2}) \footnote{
The theory with $\cos\phi=0$ has a superpotential $W=0$ when  the masses $m_i$ are zero.
It was argued that it corresponds to a non trivial fixed point \cite{kss}.
In this case the curve that describes the fivebrane factorizes to
 $t=0,z=2 C_{N_c}$ and $z=0,t=2 C_{N_c}$. It would be interesting
to study the physics of this fivebrane configuration.}.
If we work to first order in $\phi$, $\cos\phi \sim 1$
and the only nontrivial modification is that $w \sim \tilde{v} \phi$.
This is an example of one of the functions $w$ found in (\ref{wdef}),
namely the trivial case where $H(v)=0$ and $N(v)=\phi v$. 

This correspondence is very suggestive and suggests that in general
the case with $H(v)=0$ and $N(v)$ some polynomial should correspond
to $N=1$ deformations with superpotentials $\tr (\tilde{Q} \Phi^l Q)$.
However, it is also clear that in order to go beyond infinitesimal
deformations we need to also deform the relation between $t,z$ and $v$.
Therefore, the complete deformation is presumably described in
terms of two first-order differential equations that control
the change of $w$ and the $t,z,v$ equation as functions of some
deformation parameter. 

As a preliminary check of this idea, suppose the deformed brane is
simply given by replacing $v$ and $w$ by functions $\tilde{v}(v,w)$
and $\tilde{w}(v,w)$ in the second and third equation in 
(\ref{u1}). The change of variables should be holomorphic and preserve
the complex structure
\be d\tilde{v} \wedge d\tilde{w} = dv \wedge dw. \ee
We can use the third equation $\tilde{w}(v,w)=0$ to solve for
$w$ as a function of $v$, $w=w(v)$. Substituting this in the second
equation and denoting $h(v) = \tilde{v}(v,w(v))$ we find for the
deformed geometry
\bea \label{u2} \nonumber
yz & = & \Lambda_{N=2}^{2N_c-N_f} \prod_{i=1}^{N_f} (v+m_i), \\{}
\nonumber
y+z & = & 2 C_{N_c}(h(v)), \\{}
\tilde{w} & = & 0 
\stop
\eea
If $h$ has a left inverse $p(h(v))=v$, we can introduce a new variable
$v'=h(v)$, and extract from (\ref{u2}) the curve
\be \label{u3}
 t^2 - 2 C_{N_c}(v') t + \Lambda_{N=2}^{2N_c-N_f} \det(m+p(v')) =0
 \comma
\ee
which is of the form (\ref{cur2}). It would be interesting to
derive the form of $\tilde{v},\tilde{w}$ that gives rise to
a specific polynomial $p(v')$ from first principles.

Taking the curve in (\ref{u2}) as a starting point, functions $\tilde{w}$
on it are again classified by (\ref{wdef}), and one expects
that suitable functions $\tilde{w}$ correspond to adding superpotential
terms $\mu_k \tr(\Phi^k)$ as in the pure $N=2$ case. In field theory,
adding superpotentials of the form (\ref{cur2}) and $\mu_k \tr(\Phi^k)$
leads to meson vevs which are given by (\ref{bb7}) but with
$a_i + m$ replaced by $p(a_i) +m=m+\sum h_l a_i^l$. The same result
can be derived from the brane if the function $H(v)$ entering $\tilde{w}$
is given by
\be
H(v) = \prod_{j} \left( \frac{p(v)-p_j}{v-h(p_j)} \right) h'(v) \phi(h(v))
\ee
and $N(v)$ is defined by (cf. (\ref{bb11})
\be
C(v) H(v) = G_0(v) + N(v) \prod_{i=1}^l (v-p_i)
\ee
where $G_0(v)$ is a polynomial of degree $l-1$. We leave a further study
of these deformations to a future work.

\section{Superconformal Field Theory}

As is well known, $N=2$ field theories have $N=2$ critical points
where mutually non-local degrees of freedom become massless. The
simplest case is pure $SU(3)$ gauge theory which has two points
in its moduli space where mutually non-local degrees of freedom
become massless \cite{ad}. More general cases have been
studied in \cite{apsw,ehiy,eh}.
At all critical points certain dyons become massless, and these 
theories therefore can be broken to $N=1$ by suitable superpotentials
$\sum\mu_k \tr(\Phi^k)$. For $SU(3)$, it was argued in \cite{ad}
that the corresponding $N=1$ theories might correspond to
non-trivial $N=1$ fixed
points, for the points of highest criticality in $SU(N_c)$ see
\cite{f1}. Here we will rederive these arguments
from the brane picture, and study the brane configurations that
describe $N=1$ fixed points.

To show why the perturbed $N=2$ theories are candidates for
$N=1$ fixed points, we need to analyze the dyon vevs (\ref{vev}).
In the neighborhood of an $N=2$
critical point, there is a dyon condensate, and the $U(1)$ under
which the dyon is charged has a gap. For a non-trivial $N=1$
fixed point, a necessary but not sufficient condition is that
this gap has to vanish as we approach the critical
point.

As $n$ of the $q_m$ approach the $p_i$, say $q_m=p_i+\epsilon$
for $m=1,\ldots,n$ while keeping $H$ fixed, we find that 
the dyon vev $m_i\tilde{m_i}$  (\ref{vev}) behaves as $\epsilon^{n/4}$, and this
indeed vanishes as $\epsilon \rightarrow 0$. 
Geometrically, this is not a double point, and it is not completely
resolved.

Let us now study the brane configuration at $N=1$ critical 
points in more detail. Near a point where a dyon becomes
massless the brane geometry is described by
\bea
\label{p5}
(w-N(v))^2 & = &  T(v) H(v)^2
\\{}
(t-C_{N_c}(v)) H(v) &  = &  (w-N(v)) S(v).
\eea
We assume here that $H(v)$ is nonzero at
the point, say $v=p_i$, where the dyon becomes massless. If
$H(v)$ has a first order zero at this point, we can
divide the second relation by $(v-p_i)$ and still
have a good description. If $H(v)$ has a second order 
or higher order zero the two equations do not describe
the brane geometry near $v=p_i$.
Notice that the first equation in (\ref{p5}) cannot be
replaced by the equation for the $N=2$ curve, because
then the second equation at $v=p_i$ would then
reduce to $0=0$ and not constrain $w$, and this is
incorrect. 
Although the $N=2$ curve was singular at $v=p_i$, the
two equations in (\ref{p5}) describe a smooth brane
geometry, indicating that there is no longer any
massless matter at the point $v=p_i$, and indeed there
is a dyon condensate in the $N=1$ theory.

Now consider the case where 
\be
C_{N_c}(v)^2 - \Lambda_{N=2}^{2N_c-N_f} \det(v+m) = (v-p)^{2k} \prod_{i=1}^l 
(v-p_i)^2 T(v) \comma
\ee
where we distinguish two cases, either $T(v)$ has a single zero at $v=p$ or
it is nonzero at $v=p$. The deformed brane configuration is still given 
by (\ref{p5}), where $H(v)$ is of order $k+l-1$. The matrix of
derivatives of (\ref{p5}) with respect to $(w,v,t)$ at reads
\be
J = \left( \begin{array}{ccc} 2(w-N(v)) & 2(w-N)N'(v)-(TH^2)'(v) & 0 \\
-S(v) & tH'(v) -( C_{N_c}H)'(v) - 
    w S'(v) + (N S)'(v) & H(v) \end{array} \right).
\ee
One readily verifies that this matrix has rank two at $v=p$,
$t=C_{N_c}(p)$ and $w=N(p) \pm H(p) \sqrt{T(p)}$, unless
$H(p)=0$. It does not matter whether $T(v)$ is zero
at $v=p$ or not. This implies for
example that  the Argyres-Douglas points in $SU(3)$ pure
gauge theory, perturbed by a $\tr(\Phi^3)$
perturbation which corresponds to a case where $k=1$, $T(p)=0$
and $H={\rm const}$, are described
by a smooth brane configuration
\footnote{Of course, the singular point $t=0$ is infinitely far away, but
even if we include the point $t=0$ 
the brane configuration is still smooth.}. It is quite intriguing
that smooth brane configuration can give rise to
an $N=1$ fixed point. One could consider this as an
indication
that the Argyres-Douglas points perturbed by a
$\tr(\Phi^3)$ superpotential do in fact correspond to 
conventional field theories. However, as we will discuss
shortly, one cannot necessarily draw conclusions based
on local properties of the five-brane alone. 

We briefly indicate the nature of the singularity in the brane
in the singular case where $H(p)=0$. We write
that $H(v) = (v-p)^r \tilde{H}(v)$, and assume that $r \leq k$.
We will also decompose $T(v)=(v-p)^{\nu} \tilde{T}(v)$ with
$\nu=\pm 1$, and $S(v)=(v-p)^{k} \tilde{S}(v)$.
In terms of the new variables $\tilde{v}=v-p$,
$\tilde{w} = w-N(v)$, and $y=t-C_{N_c}(v)$,
the geometry reads
\bea
\tilde{w}^2 & = & \tilde{v}^{2r+\nu} \tilde{T}(\tilde{v}) \tilde{H}(\tilde{v})^2 \\{}
y \tilde{H}(\tilde{v}) & = & \tilde{v}^{k-r} \tilde{S}(\tilde{v}) \tilde{w}
\eea
and near $v=p$ we effectively have
\be \tilde{w}^2  =  \tilde{v}^{2r+\nu}, \qquad y=\tilde{v}^{k-r} \tilde{w}.
\label{sq1}
\ee
This clearly describes a singular brane configuration. In order to determine
whether this brane configuration corresponds to a non-trivial $N=1$ fixed point,
we would need to determine the dimensions of some of the operators at
the fixed point. In the next section we will discuss how
this can in principle be done using
the fivebrane configuration. A detailed calculation will be left for future work.

\subsection{Scaling Dimensions, the K\"ahler Potential and the Superpotential}

At the $N=1$ fixed point, there is an unbroken $U(1)_R$ symmetry. Unfortunately,
it is an accidental symmetry, making it hard to determine the dimensions of the
chiral fields that give rise to relevant perturbations. 
In $N=2$ theories, the dimensions can be determined using the fact that contour
integrals of the Seiberg-Witten differential needs to have dimension one 
\cite{ad,apsw,ehiy,eh}. In the $N=1$ theories discussed in \cite{aks},
the holomorphic three-form of Calabi-Yau manifolds could be used
to fix the dimensions of operators. What distinguishes these two cases from 
the present one is that in both the dimensions could be determined by a
local calculation near the singularity in the Riemann surface and the
Calabi-Yau manifold respectively. In the present case, the complete global
structure of the fivebrane seems relevant in order
to determine the normalization of fields,
and this explains why smooth brane configurations can still correspond to
non-trivial $N=1$ theories. 

If there is a physical quantity with a fixed dimension
which depends only on a neighborhood
of the point $\tilde{v}=\tilde{w}=\tilde{t}=0$, then we can read of the
relative dimensions of $\tilde{v},\tilde{w}$ and $\tilde{t}$
from (\ref{sq1}) using the obvious $U(1)_R$ symmetry it possesses
(and using the fact that $\Lambda_{N=2}$ has $U(1)_R$
weight zero). The overall normalization of operators would then
follow from the scaling behavior of this physical 
quantity\footnote{At the point of highest criticality in pure $SU(N_c)$
gauge theory with the most singular choice of $H$ in (\ref{sq1}),
we have $2r+\nu=N_c-2$ and $k-r=1$. Then (\ref{sq1}) would yield
the following relations between dimensions (indicated by square 
brackets): $2[\tilde{w}]=2[\tilde{H}] + (N_c-2) [\tilde{v}]$ and
$[y]+[\tilde{H}] = [\tilde{w}] + [\tilde{v}]$. In addition
the superpotential contains $\tilde{H} \tr(\Phi^{N_c})$ leading
to an additional relation $[\tilde{H}]+N_c[\tilde{v}]=3$. 
From this one can derive that $[y]=\frac{N_c(3-[\tilde{w}])}{
N_c+2}$ and $[\tilde{v}]=\frac{2(3-[\tilde{w}])}{
N_c+2}$. At the $N=2$ point the dimensions of $\tilde{v},y$
\cite{ehiy} are recovered if we take $[\tilde{w}]=2$, the
standard dimension of a meson in the UV. If for some reason
the dimension of $\tilde{w}$ would remain two even in the perturbed
theory, the dimensions of operators in the $N=1$ points would be
the same as those at the $N=2$ points.}.

However, if there is no physical quantity which depends only a neighborhood of 
the point $\tilde{v}=\tilde{w}=\tilde{t}=0$, then
in order to determine the dimensions of operators, we need to explicitly compute
another object whose dimension is known. Examples are the K\"ahler potential, which
has dimension two, and the superpotential, which has dimension three. 
The K\"ahler potential is a non-holomorphic quantity
and its computation using the brane geometry is still an open problem.
The complication in performing this computation is the need for 
a consistent decoupling of the Kaluza-Klein modes.

The computation of the superpotential seems an easier task. In
\cite{wittennew} the following expression for the superpotential
was proposed. Let $\Sigma\subset {\bf R}^5 \times S^1$ 
be the Riemann surface part of the fivebrane, and $\Sigma_0
\subset {\bf R}^5 \times S^1$ be another Riemann surface
which has the same asymptotic behavior at infinity. If 
$B\subset {\bf R}^5 \times S^1$ is a three-manifold with
boundary $\Sigma-\Sigma_0$, then 
\be W(\Sigma)-W(\Sigma_0) = \int_B \Omega
\ee
where $\Omega$ is the holomorphic three-form
\be
\Omega=R \frac{dt}{t} \wedge dv \wedge dw.
\ee
This superpotential has been computed in some examples
in \cite{wittennew,21}.
One of the problems with this definition of the superpotential is
the dependence on the the surface $\Sigma_0$. We are only interested in
the behavior of the superpotential as a function of some parameter,
say $g$. If we change a parameter it is possible that we change the
asymptotic behavior of the fivebrane, in which case we would
have to choose a new surface $\Sigma_0$ as well. The dependence
of $\Sigma_0$ on $g$ is not fixed by anything, making the outcome
of this calculation highly ambiguous. Keeping $\Sigma_0$ fixed anyway
would lead to the result
\be \label{rr0}
\frac{\partial W}{\partial g} = \int_{\Sigma} \Omega_2
\ee
with 
\be \label{rr1}
\Omega_2 = \frac{R}{t} \left( \frac{\partial t}{\partial g} dv \wedge dw + 
\frac{\partial v}{\partial g} dw \wedge dt +
\frac{\partial w}{\partial g} dt \wedge dv \right).
\ee
On the other hand, in all examples considered in \cite{wittennew,21}
the superpotential turns out to be, roughly speaking,
a weighted sum of contour integrals
\be \label{rr2}
W \sim \sum a_iR \oint_{C_i} vw\frac{dt}{t}
\ee 
where the $C_i$ are contour integrals around the various infinite `tubes'
that stick out of the Riemann surface $\Sigma$. A
reduction of the expression in (\ref{rr1}) to one of the type (\ref{rr2})
will be useful in order to arrive at an expression for the superpotential
which does correctly reproduce the field theory superpotentials without
having to choose an additional surface $\Sigma_0$. With such an
expression it would be quite easy to determine for example the
dimensions of operators at an $N=1$ critical point. Notice that
(\ref{rr2}) does depend on the asymptotic behavior of the brane,
and does not seem to care whether the brane has a singular point
or not.
We will not carry out the computation here but comment
  how we can recognize several pieces of the
field theory superpotential in (\ref{rr2}). As $v\rightarrow -m_i$ and
$w\rightarrow w_i$, there is an infinite tube stretching out with
$t \rightarrow 0$. The contour integral around this tube contributes
in (\ref{rr2}) a term proportional to $m_i w_i$. Here we recognize
the term in the superpotential which is simply the mass term for the meson
$M_i$. This provides another explanation why
the finite values of $w$ as $v\rightarrow -m_i$ should be related to 
the  meson vevs. 

Another term can be seen in the case where we have the maximal number
of mutually massless dyons, in the parametrization (\ref{au60}) and
(\ref{wsol}). Then there is an infinite tube with $q\rightarrow 0$,
and the contour integral around that tube yields a term proportional
to $q_+ q_- w_1$, where $w_1$ is the coefficient of the term in
$w$ that is linear in $q$ in (\ref{wsol}). It is a highly non-trivial
result that this is proportional to $\sum_k \mu_k u_k$, where
$u_k$ is the vev of $\tr(\Phi^k)$ at the point with the maximal 
number of massless dyons. The explicit form of $u_k$ is given
in (\ref{au62}). Thus it seems that a suitable version of (\ref{rr2})
does reproduce precisely the field theory superpotential.

\section{Example: $SU(3)$ with $N_f=2$}

To illustrate several of the things discussed in the paper, we will now discuss the example
of $SU(3)$ gauge theory with two flavors and superpotential $W=\mu_2 \tr(\Phi^2) + 
\mu_3 \tr(\Phi^3)$ in some detail. The $N=2$ theory is described by the curve
\be
t^2 - 2 t (v^3 - \frac{u_2}{2} v - \frac{u_3}{3}) + \Lambda_{N=2}^4 (v+m_1)(v+m_2) =0
\ee
The first thing is to find the locus in moduli space where a dyon becomes massless.
At this locus, we have
\be
 (v^3 - \frac{u_2}{2} v - \frac{u_3}{3})^2 -  \Lambda_{N=2}^4 (v+m_1)(v+m_2)
=(v-a)^2 T(v) 
\ee
for some $a$. From this we find that $u_2,u_3$ should be given by
\bea u_2 & = & 6 a^2 - b^{-1} (2a+m_1+m_2) \Lambda_{N=2}^2
\nonumber  \\{}
u_3 & = & -6a^3 + \frac{3}{2} a b^{-1} (2a+m_1+m_2) \Lambda_{N=2}^2
-3b\Lambda_{N=2}^2
\label{ex0}
\eea
where
\be b=\pm \sqrt{(a+m_1)(a+m_2)}.
\ee
We will take the plus sign in $b$ from now on, the discussion with a minus
sign in $b$ proceeds in a similar way. 
Now recall that the deformed brane configuration is described by
(see (\ref{wdef}) and (\ref{au12}), $H$ must in this case be
equal to a constant)
\bea
w^2 - 2 N(v) w + N(v)^2 - H^2 T(v) & = & 0 \\{}
(w-N(v)) (v-a) & = & (t-C_3(v)) H
\eea
and by explicit computation we find
\bea \label{ex1}
N(v) & = & H (v^2 + a v + (a^2 - \frac{u_2}{2})) \\{}
N^2-H^2 T & = & H^2(-2a^3 + au_2 + \frac{2}{3} u_3) v + 
  H^2 (-4a^4 + \Lambda_{N=2}^4 + 2 a^2 u_2 + \frac{4}{3} a u_3)
\eea
For the dyon vev we find
\be
(m\tilde{m} )^2 = H^2 \sqrt{T(a)} = H^2(6a \Lambda_{N=2}^2 b 
+ \frac{(m_1 + m_2)^2}{4 b^2} \Lambda_{N=2}^4).
\ee
The asymptotic behavior of $w$ tells us which values of $\mu_2,\mu_3$
the deformed brane corresponds to. From (\ref{ex1}) we see that
$w\sim 2Hv^2 + 2Ha v$ and therefore
\be 3\mu_3 = 2H, \qquad  2\mu_2 =2Ha
\ee
consistent with what we found in general in (\ref{518}).

Using (\ref{ex0}) and (\ref{ex1}) we can solve explicitly for
$w$ as $v=-m_1$ or $v=-m_2$. We find that in both cases
one of the two solutions is equal to 
\be w_i = H\Lambda_{N=2}^2 \frac{b}{a+m_i}
\ee
which confirms (\ref{Wi}).

Let us now consider what happens when we tune the parameters further,
so that we get more massless particles.
On the one hand, we can tune parameters so that mutually non-local
dyons become massless. This happens whenever $T(v)$ will be divisible
by additional factors of $(v-a)$. The point of highest criticality is
reached when 
\be
u_2 = 15a^2,\quad u_3=-60a^3,\quad m_1+m_2=-\frac{16}{5} a ,\quad m_1 m_2 = 4 a^2
\ee
and
\be a^4 = \frac{4}{405} \Lambda^4.
\ee
For these values of the parameters the equation for the $N=2$ curve reads
\be
y^2 = (v-a)^5(v+5 a)
\ee
where $y= t-C_3(v)$. One way to approach this point of highest
criticality is to start with a point with two mutually local massless dyons where
$y^2 = (v-a)^2(v-b)^2 T(v)$, and then to tune the parameters
in such a way that $b\rightarrow a$ and $T(v)$ becomes divisible by $(v-a)$. 
When there are two mutually local massless dyons, supersymmetry is
unbroken for each value of $\mu_2$ and $\mu_3$ in the superpotential,
and this shows that the same is true for the point of highest criticality. 
The equations describing the brane configuration at this point read
(where $\tilde{w}=w-N(v)$ and $\tilde{v}=v-a$)
\be
\tilde{w}^2 = H(\tilde{v})^2 \tilde{v}(\tilde{v}+6a), \qquad 
y H(\tilde{v}) = \tilde{v}^2 \tilde{w}.
\ee
This brane configuration is smooth unless $H(\tilde{v})\sim\tilde{v}$,
in which case it reduces to (near $\tilde{v}=0$)
$\tilde{w}^2 =6a\tilde{v}^3$, $y=\tilde{v}\tilde{w}$.
To reach this singular brane configuration we need to take
the parameters in the superpotential such that $2\mu_2 = 3a\mu_3$.
We expect therefore a qualitative difference between the different $N=1$
theories obtained by perturbing the highest $N=2$ critical point, 
depending on whether $2\mu_2 = 3a\mu_3$ or not. One could speculatethat
only in the latter case we obtain an $N=1$ superconformal field
theory, but for this we need a more detailed understanding of the
relation between brane geomety and the appearance of 
superconformal fixed points in field theory, as discussed in the
previous section.

Finally, we consider the points where two mutually local dyons become
massless. We take $m_1=m_2=0$ for simplicity. There are two
distinct possibilities, either
\be
C_3(v)^2 - \Lambda_{N=2}^4 v^2 = v^4 (v^2 \pm 2 \Lambda^2) 
\ee
with 
\be u_2=\mp 2\Lambda^2
\ee
or 
\be
C_3(v)^2 - \Lambda_{N=2}^4 v^2 = (v-a)^2 (v-\omega a)^2 (v+2a)(v+2\omega a)
\ee
with
\be
\omega^3=1, \quad a=\frac{2}{9} \Lambda^2 (1-\omega), \quad
u_2 = 2(3a^2 -\Lambda^2), \quad u_3=-6a^3.
\ee
These two cases correspond to the non-baryonic branch roots with $r=1,0$ 
respectively. We take the first case as our example, and take $H(v)=h_0 v + h_1$.
Then $N(v)=h_0 v^2 + h_1 v \pm h_0 \Lambda_{N=2}^2$, and we see that
$3 \mu_3 = h_0$ and $2\mu_2 = h_1$. The equation for $w$ reads
\be
w^2 - 2 N(v) w - 2 h_0 h_1 \Lambda_{N=2}^2 v + 
 h_0^2 \Lambda_{N=2}^4 \mp 2 h_1^2 \Lambda_{N=2}^2=0.
\ee
If we now make the substitution 
\be v=q \mp \frac{1}{2} \Lambda_{N=2}^2 q^{-1}
\ee
then the equation for $w$ factorizes,
\be
(w-2h_0 q^2 - 2 h_1 q)(w \pm h_1 \Lambda_{N=2}^2 q^{-1} - 
\frac{1}{2} h_0 \Lambda_{N=2}^4 q^{-2} )=0
\ee
and so does the equation for $t$,
\be
\frac{1}{4q^3} (t-2q^3 \pm \Lambda^2 q)(\pm \Lambda^6 - 2 \Lambda^4 q^2 + 4 q^3 t)=0.
\ee
The equation for the deformed brane is obtained by taking both for $w$ and $t$ the
first factor in the equations. In particular, $w=2h_0 q^2 + 2 h_1 q$, 
in perfect agreement with (\ref{wsol}) and (\ref{au60}). The parameters
$q_+$ and $q_-$ are given by $q_+=\sqrt{\pm\frac{1}{2}} \Lambda_{N=2}$,
$q_-=-q_+$. The meson vevs are simply given by substituting
$q_{\pm}$ in the expression for $w$. The results agree with (\ref{au50}),
which tells us that $w_{\pm} = 2 h_1 q_{\pm} -2 h_0 (q_{\pm}^2 + 2 q_{\pm} q_{\mp})$,
upon using $q_-=-q_+$. 

Finally, we compare these meson vevs with the ones obtained from (\ref{bb12}). 
With two massless dyons, (\ref{bb12}) reduces
to
\be
w_j= \frac{f_j(p_1)-f_j(p_2)}{(p_2-p_1) } 
\Lambda_{N=2}^2,
\ee 
where
\be
f_j(p) = \frac{H(p) \det(m+p)^{1/2}}{p+m_j}.
\ee
As we take $p_1 \rightarrow p_2$, $w_j$ approaches
$-f_j'(p_1)$, but this does not have a 
well-defined limit as we send $p\rightarrow 0$ and 
$m_i \rightarrow 0$. This 
problem can be traced back to the derivation of (\ref{bb12}),
where we assumed there that $m_j+p_i$ is not equal to zero.
If $p_i=m_{2i-1}=m_{2i}$ for $i=1,\ldots,r$, then $C_{N_c}(v)$ must
also be divisible by $\prod_{i=1}^r(v-p_i)$,
and (\ref{bb10}) becomes
\be
\tilde{C}_{N_c-r}(v)^2 - 
\Lambda_{N=2}^{2N_c-N_f} \prod_{i=2r+1}^{N_f} (v+m_i) =
\prod_{i=r+1}^l (v-p_i)^2 T(v)
\ee
The derivation of (\ref{bb12}) remains the same, except
that in the final answer only the masses
$m_j$ with $j>2r$ and the $p_i$ with $i>r$
appear. Equation (\ref{bb12}) therefore
only provides us with the meson vevs of
the mesons $M_j$ with $j>2r$. The other meson
vevs $M_j$ with $j\leq 2r$ cannot be written
in a simple way. This also explains why when
we consider a theory at a NB branch root,
we only get at most one meson vev from (\ref{bb12}).

Coming back to the example we are considering, 
after taking out a factor of $v^2$ from (\ref{bb10}),
we are left with no masses at all, and in this case
(\ref{bb12}) does unfortunately not yield any
information about the meson vevs. This happens 
always at the NB branch root with $2r=N_f$.

\section{Discussion}

The present work points at several interesting directions to pursue further.
We have shown, using the results obtained from the
fivebrane, that while for vacua where the classical enhanced gauge group 
is $SU(2)$ the effective superpotential 
obtained by the "integrating in" method is exact,
this is no longer true when the classical enhanced gauge group is $SU(r), r >2$.
It would be interesting to find $W_{\Delta}$ in this case, and to see
which singular submanifolds of the 
$N=2$ Coulomb branch the superpotential parametrizes. 
Finding $W_{\Delta}$ can presumably be done by
computing the superpotential using the fivebrane 
configuration as suggested by \cite{wittennew}.

We have derived several results from the fivebrane which should be understood from 
the field theory viewpoint.
Among these phenomena are the splitting of the baryonic branch in section 5
and the simple expression for the meson vev (\ref{au50}).
It is also curious to note that at 
the roots
of the non-baryonic branch with a maximal number of massless
dyons, there
exists an expression for the gauge invariant order parameters
$u_k$ of the form  
\be \label{au62}
u_k = (-1)^k (k-1) (2N_c-N_f) q_+ q_- \sum_{l=0}^{k-2}
q_+^l q_-^{k-2-l} \left( \begin{array}{c} k-1 \\ l 
\end{array} \right) \left( \begin{array}{c} k-2 \\ l
\end{array} \right) \frac{1}{l+1} 
\comma
\ee
and the finite values of $w$ (\ref{au50}) are proportional
to the derivatives of $u_k$ with respect to $q_{\pm}$.
The field theory explanation for this should presumably
rely on identifying an appropriate set of operators to describe
the modified non-baryonic branches in view of the perturbation 
$\Delta W$ (\ref{dw}).
These are all parts of a bigger picture describing
the complete structure of the Higgs branches
and their intersection with the Coulomb branch
when we perturb the $N=2$ gauge theory by $\Delta W$ (\ref{dw}).
We have clearly seen that this structure is 
much richer than in the case where the perturbation
includes only a mass term for the adjoint 
chiral multiplet and it deserves further study. 

Related to the above mentioned problem of identifying the
appropriate set of operators in each situation are the
various dualities the field theory can possess. The brane
gives a uniform geometrical description, whereas the
field theoretical descriptions depend on the relevant 
weakly coupled degrees of freedom. The field
theory has has for example
a dual magnetic description \cite{kut,ks}, and
as in the case of pure $N=1$ gauge theory with
matter this may be related to interchanging
the role of $v$ and $w$ \cite{23}. On the other
hand, (\ref{au12}) looks like a curve for a
$SU(N_c-l)$ gauge theory with $N_c-l-1$ flavors,
so by interchanging the role of $t$ and $w$ we
may end up with such a completely novel
dual description. 

Other $N=2$ theories like the product gauge theories in
\cite{witten} can also be perturbed by superpotentials
in the brane framework. If there are only mass terms
for the adjoints, the relevant configuration is presumably
the one given in \cite{21}. Perturbing these theories
by higher order superpotentials can lead to new families
of $N=1$ fixed points. 

For fixed $\mu_k$ there are generally only a finite
number of points on the Coulomb branch that are not
lifted. Thus, there can be domain walls as in
\cite{wittennew} and it would be interesting to know
their behavior as a function of $\mu_k$.

We have not completed the study of
the $N=1$ gauge theories with Landau-Ginzburg type 
superpotential  in the fivebrane
framework. This
is an interesting direction by itself which should be pursued in order to
learn about field theory from the fivebrane.  
 
The study  of non trivial IR fixed points 
using the M theory  fivebrane
is also not completed. In particular it
is still not clear whether the fivebrane
can geometrically distinguish between trivial
and not trivial fixed points.
We have not completed the calculation of 
the superpotential at the fixed points.
A related issue is to study the case with 
vanishing superpotential \cite{kut,ks,kss}
on which we commented
in section 7.

\noindent
{\bf Acknowledgements}

We would like to thank M. Douglas, A. Giveon,
K. Hori, S. Kachru, V. Kaplunovsky,  H. Ooguri, O. Pelc, M. Peskin and 
J. Terning for valuable 
discussions.  JdB and YO would
like to thank the theory division at CERN for hospitality.
This research is supported in part by 
NSF grant PHY-95-14797 and DOE grant DE-AC03-76SF00098.
JdB is a fellow of the Miller Institute for Basic Research in
Science.

\end{document}